\documentclass[sigconf,nonacm]{aamas} 
\usepackage{amsmath}
\usepackage{amsthm}
\usepackage{xfrac}
\usepackage[noend]{algorithmic}
\usepackage{algorithm}
\usepackage{mathtools}
\usepackage{subcaption}
\usepackage{pgfplots}
\usepackage{filecontents} 
\usepackage{enumitem} 
\usepackage{adjustbox}
\usepackage{arydshln} 

\usetikzlibrary{bayesnet}
\usetikzlibrary{calc,arrows.meta,positioning}

\newcommand{\states}{\mathcal{S}}
\newcommand{\actions}{\mathcal{A}}

\newcommand{\transfunc}{\mathcal{T}}

\newcommand{\mg}{\mathcal{G}}

\newcommand{\ba}{\mathbf{a}}
\newcommand{\bpi}{{\boldsymbol{\pi}}}
\newcommand{\bl}{\boldsymbol{\lambda}}
\newcommand{\br}{\mathbf{r}}
\newcommand{\bR}{\mathbf{R}}

\newcommand{\equilibriums}{{\mathcal{D}}}
\newcommand{\bd}{{\boldsymbol{d}}}
\newcommand{\demos}{\mathcal{D}}
\newcommand{\demosall}{\mathcal{X}}
\newcommand{\Rset}{\mathcal{R}}

\newcommand{\ent}{\mathcal{H}}
\newcommand{\group}{\mathbf{g}}

\newcommand{\bel}{\mathbb{P}}

\newcommand{\pool}{\mathcal{P}}

\newcommand{\dkl}{D_\text{KL}}
\newcommand{\gapv}{\Delta^\text{QIG}}
\newcommand{\gapkl}{\Delta^\text{PSG}}

\newcommand{\approptoinn}[2]{\mathrel{\vcenter{
  \offinterlineskip\halign{\hfil$##$\cr
    #1\propto\cr\noalign{\kern2pt}#1\sim\cr\noalign{\kern-2pt}}}}}

\newcommand{\appropto}{\mathpalette\approptoinn\relax}

\newtheorem{assumption}{Assumption}[section]
\newtheorem{definition}{Definition}

\newtheorem{example}{Example}
\newtheorem{theorem}{Theorem}
\newtheorem{corollary}{Corollary}
\newtheorem{proposition}{Proposition}

\usepackage{balance} 

\setcopyright{ifaamas}
\acmConference[AAMAS '26]{Proc.\@ of the 25th International Conference
on Autonomous Agents and Multiagent Systems (AAMAS 2026)}{May 25 -- 29, 2026}
{Paphos, Cyprus}{C.~Amato, L.~Dennis, V.~Mascardi, J.~Thangarajah (eds.)}
\copyrightyear{2026}
\acmYear{2026}
\acmDOI{}
\acmPrice{}
\acmISBN{}

\acmSubmissionID{314}

\title[Inference of Altruism and Intrinsic Rewards in Multi-Agent Systems]{Inference of Altruism and Intrinsic Rewards in Multi-Agent Systems}

\author{Victor Villin}
\affiliation{
  \institution{University of Neuchâtel}
  \city{Neuchâtel}
  \country{Switzerland}}
\email{victor.villin@unine.ch}

\author{Christos Dimitrakakis}
\affiliation{
  \institution{University of Neuchâtel}
  \city{Neuchâtel}
  \country{Switzerland}}
\email{victor.villin@unine.ch}

\begin{abstract}

Human interactions are influenced by emotions, temperament, and affection, often conflicting with individuals' underlying preferences.
Without explicit knowledge of those preferences, judging whether behaviour is appropriate becomes guesswork, leaving us highly prone to misinterpretation.
Yet, such understanding is critical if autonomous agents are to collaborate effectively with humans. We frame the problem with multi-agent inverse reinforcement learning and show that even a simple model, where agents weigh their own welfare against that of others, can cover a wide range of social behaviours.
Using novel Bayesian techniques, we find that intrinsic rewards and altruistic tendencies can be reliably identified by placing agents in different groups.
Crucially, this disentanglement of intrinsic motivation from altruism enables the synthesis of new behaviours aligned with any desired level of altruism, even when demonstrations are drawn from restricted behaviour profiles.

\end{abstract}


\newcommand{\BibTeX}{\rm B\kern-.05em{\sc i\kern-.025em b}\kern-.08em\TeX}

\begin{document}


\pagestyle{fancy}
\fancyhead{}


\maketitle 


\section{Introduction}

Multi-Agent Inverse Reinforcement Learning (MAIRL) seeks to uncover the hidden reward structures that govern interacting agents. By inferring these latent motivations, we can either interpret observed behaviours~\citep{chen2024unveiling} or train policies that align with them~\citep{hadfield2016cooperative, yu2019multi, haynam2025multi}. Yet, human interactions are rarely straightforward. Assuming that an agent's actions directly reflect their personal rewards is dangerously simplistic: the same behaviour could stem from a desire to help, harm, or manipulate others. Misreading these intentions can lead to fundamentally flawed models of social behaviour.

As Artificial Intelligence (AI) systems become increasingly pervasive, 
aligning them with human values is no longer optional~\citep{dung2023current}. AI must not only infer agents' preferences but also evaluate whether these preferences are socially beneficial. Detecting harmful or counterproductive behaviours is essential for designing policies that reliably elevate human welfare.

Reward inference is notoriously challenging in multi-agent systems~\citep{freihaut2025feasiblerewardsmultiagentinverse}. Classical approaches assume zero-sum or fully cooperative structures~\citep{lin2019multi}. Real-world interactions, however, are far more nuanced: humans are neither perfectly competitive nor purely cooperative. A chess player may deliberately hold back while coworkers may occasionally act selfishly, deviating from collective benefit. Interactions in society are mostly general-sum.

Among the factors that shape observed preferences, \emph{altruism} plays a central role. Psychologically, altruism is defined as `any behavior that increases another person's welfare'~\citep{ellen1993altruism, batson2014altruism}. The \emph{altruism scale}~\citep{sawyer1966altruism} frames this as a continuum: negative altruism corresponds to antagonistic behaviour, zero to pure self-interest, and positive altruism to prosociality. More precisely, \citet{sawyer1966altruism} interprets altruism as the weight an individual places on the welfare of others relative to their own.

Motivated by this perspective, we study the case where rewards of an agent are a linear combination of its intrinsic rewards and those of other, weighted by a latent altruism level. This gives rise to a two-fold inference problem: (1) recovering the intrinsic rewards that drive each agent's behaviour, and (2) estimating how each agent balances self-interest with concern for others.  

Understanding altruism is crucial for interpreting and guiding social behaviour. It explains deviations from cooperative goals, supports team management to encourage prosociality, and helps diagnose potentially misaligned AI agents. We show that disentangling altruism from rewards enables the design of agents that are consistently altruistic, by acting according to the inferred intrinsic rewards of others. Ultimately, this leads to AI systems that are not only effective, but also interpretable, trustworthy, and socially aware.

\paragraph{Contributions.} In summary, our main contributions are:
\begin{enumerate}[topsep=4pt, leftmargin=12pt]
\item We formally introduce MAIRL with altruism-structured rewards (Section~\ref{sec:setting}), forming a general-sum problem constrained by the fact that agents rewards are modeled as linear combinations of their intrinsic rewards and those of others.
\item We analyse identifiability under the altruism scale model (Section~\ref{sec:identifiability}) and reveal that rewards are generally hard to recover. We prove that leveraging demonstrations from multiple agent groups closes this identifiability gap. Notably, we uncover similar results to Inverse Reinforcement Learning (IRL) about identifiability \citep{rolland2022identifiability}.
\item We propose two novel Bayesian methods for learning from multi-group demonstrations (Section~\ref{sec:inference}). The first adapts Bayesian inverse reinforcement learning to the multi-agent setting by constructing a reward posterior assuming Boltzmann rationality. The second, our main contribution, does
 \emph{not} require rationality assumption: it first infers a policy posterior from demonstrations, then derives a reward posterior conditioned on the policy posterior.
\item We conduct extensive experiments to validate our approaches on challenging sets of random Markov Games (Section~\ref{sec:experiments}). We further consider a practical Overcooked scenario~\citep{carroll_utility_learning_about_2019}, where anti-social chefs must collaborate, and test whether we can synthesise behaviours at new altruism levels. We compare against other state-of-the-art MAIRL techniques, namely Multi-Agent Marginal Q-Learning (MAMQL~\citep{haynam2025multi}) and Multi-Agent Adversarial IRL (MAAIRL~\citep{yu2019multi}).
\item We show that our approach can accurately recover both intrinsic rewards and altruism levels. Most importantly, we demonstrate that synthesising altruistic behaviour from entangled estimates can produce adversarial policies. In contrast, by leveraging demonstrations from multiple groups, our approach identify rewards and robustly generalises to behaviours aligned with previously unseen altruism levels.
\end{enumerate}



%


\section{Related Work}

\textit{IRL and MAIRL} aim to infer reward functions that explain observed behaviours, assuming these are near-optimal~\citep{ng2000algorithms, natarajan2010multi}. Reward identification in IRL is inherently ill-posed: even under entropy regularisation, rewards are only identifiable up to potential-based shaping transformations~\citep{ng1999theory}. Recovering them up to additive constants further requires demonstrations under varying dynamics~\citep{cao2021identifiability, rolland2022identifiability, buening2024environment, schlaginhaufen2024towards}. In multi-agent settings, the notion of optimality is considerably more intricate. Rather than maximising individual rewards, agents interact strategically, leading to different formulations of MAIRL. Some works simplify the problem by decomposing it into independent single-agent IRL tasks~\citep{lin2019multi, fu2021evaluating}, while others explicitly model equilibrium concepts such as Nash equilibria~\citep{reddy2012inverse, martin2021bayesian, bergerson2021multi, mehr2023maximum}.
Despite these advances, theoretical understanding of reward identifiability in MAIRL remains limited. \citet{freihaut2025feasiblerewardsmultiagentinverse} characterised the feasible set of reward functions consistent with demonstrations but did not establish identification results. In this work, we prove that we can achieve identification under widely adopted reward assumptions~\citep{reddy2012inverse, freihaut2025feasiblerewardsmultiagentinverse}, by placing agents in different groups. This is similar to how changing a partner's policy reveals information about the rewards of others ~\citep{buning2022interactive}.

MAIRL has seen successful applications in fully cooperative and zero-sum domains~\citep{jeon2020scalable, fu2021evaluating, martin2021bayesian}, where reward structures are tightly constrained. Now, extending it to general-sum games poses new challenges. While maximum-entropy MAIRL and inverse Q-learning approaches can reproduce expert-like behaviours~\citep{yu2019multi, haynam2025multi}, their learned rewards are not guaranteed to be interpretable or socially consistent. Nevertheless, it has been successfully applied to analyse human driving habits~\citep{mehr2023maximum}, and related efforts have explored theory of mind formulations for inferring others' intentions~\citep{GToM:Yoshida:2008, chen2024unveiling}. We demonstrate however that such methods fail to learn a disentangled understanding of social preferences, and misunderstand agent intrinsic rewards.



\textit{Bayesian IRL} is a framework for inferring rewards through
probabilistic reasoning~\citep{ramachandran2007bayesian}. Despite its
success in single-agent scenarios~\citep{choi2011map,
  rothkopf2011preference,dimitrakakis2011bayesian}, its application to
MAIRL remains underexplored~\citep{lin2017multiagent,
  martin2021bayesian}.  Common drawbacks to existing MAIRL methods are
strong assumption about behaviours, such as strict rationality or a
specific amount of entropy regularisation~\citep{yu2019multi,
  lin2019multi}. \emph{We propose} a Bayesian modelling
approach that only places a prior on the optimality of the policies, which
we show performs significantly better than approaches which do make
behavioural assumptions.











\textit{Altruism} plays a fundamental role in decision-making and social interaction. The problem of inferring altruistic behaviour has been explored across both behavioural economics and psychology. In these fields, altruism is typically modeled as a linear concern for others' welfare. \citet{charness2002understanding} showed that, under such models, individuals tend to increase collective payoffs even at a personal cost. From a psychological standpoint, \citet{sawyer1966altruism} proposed representing altruism as a weight on others' welfare along a continuum ranging from negative (adversarial) to neutral (egoistic) to positive (altruistic). However, these studies are limited to controlled surveys or stylised economic games.
In multi-agent reinforcement learning, altruism has been incorporated in a similar linear form, combining an agent's own reward with those of its teammates to promote cooperation and prosocial behaviour~\citep{prosocial2018peysakhovich, hughes2018inequity, berner2019dota, hostallero2020inducing, agapiou_melting_pot_2_2023}. Related approaches based on reputation dynamics also encourage cooperative behaviour, and can be viewed as a more complex generalisation of altruism: agents prefer to act altruistically toward others with good reputations~\citep{anastassacos2021cooperation, ren2023reputation}. In MAIRL, \citet{fukumoto2020cooperative} showed that cooperative policies can be induced from selfish demonstrations by augmenting expert data with generated samples, though this approach does not explicitly model altruism.

We study the case where each agent has its own intrinsic level of altruism, independent of the individuals it interacts with. This aligns with prior approaches while extending the continuous altruism scale of \citet{sawyer1966altruism}. To the best of our knowledge, we are the first to employ MAIRL to infer altruism in dynamic multi-agent games and to generate behaviours across the full altruism spectrum.

\section{Preliminaries}
\label{sec:prelims}

\paragraph{Rewardless Markov Games.}
An $n$-player Rewardless Markov Game (RMG) can be formalised as a tuple $\mg = \mg(\cdot) = (\states, \boldsymbol{\mathcal{A}}, \transfunc, \gamma, \omega_0, \cdot)$, where $\states$ is a set of states, $\boldsymbol{\mathcal{A}} = \actions_1 \times \dots \times \actions_n$ is a set of discrete actions for each player, $\transfunc: \states \times \boldsymbol{\mathcal{A}} \rightarrow \Delta^\states$ is a transition function, $\gamma \in [0, 1[$ is a discount factor, and $\omega_0$ an initial state distribution. Without loss of generality, we focus on games where all players share the same action space, i.e. $\boldsymbol{\mathcal{A}} = \actions^n$. We label the discrete actions as $\{a^1, \dots, a^{|\actions|}\} = \actions$. For clarity, we study player-permutation invariant games, meaning dynamics remain unchanged upon player reordering (e.g. agent $A$ playing with agent $B$ is the same as $B$ playing with $A$).\footnote{Player-permutation invariant games are not restrictive. Many real-world social interactions (e.g., trading in markets, sports such as football) are inherently symmetric in dynamics. Here, no structural advantage or disadvantage arises purely from the dynamics, but rather from the agents' preferences.}

\paragraph{Policies.}
A policy $\pi: \states \rightarrow \Delta^\actions$ is a probability distribution over a single agent's actions. A joint policy is given by $\bpi = (\pi_1, \dots, \pi_n) = (\pi_i, \bpi_{-i})$, where $\bpi_{-i}$ refers to the joint policy of all policies except policy $i$.
We denote the set of policies by $\Pi$. We assume actions are conditionally independent, so that the probability of the joint action $\ba \in \actions^m$ at state $s$ under the joint policy $\bpi$ is $\bpi(\ba \mid s) = \prod_i \pi_i(a_i \mid s)$. We also denote by $\transfunc^{\bpi_{-i}}_{a}(s) = \mathbb{E}_{\ba_{-i}\sim \bpi_{-i}}\left[\transfunc(\cdot \mid s, a, \ba_{-i})\right]$ transitions of agent $i$  induced by the other agents' policies when picking action $a$, and $\transfunc^\bpi = \mathbb{E}_{\bpi}\left[\transfunc\right]$ the transition induced by the full joint policy.

\paragraph{Rewards and regularised values.}
The reward function of an agent $i$, $R_i: \states \times A \rightarrow \mathbb{R}$, outputs a bounded real value given a state and a joint action. We write the joint reward function as $\bR=(R_i)_i$. Given a reward function $\bR$, a general-sum Markov game (MG) is defined as $\mg(\bR)$. For a joint policy $\bpi$ on $\mg(\bR)$, the entropy-regularised value function of agent $i$ is
\[
    V^\bpi_i(s) \coloneqq \mathbb{E}^{\bpi}_{\mg} \left[ \sum_{t=0}^\infty \gamma^t R_i(s_t, \ba_t) + \frac{1}{\beta} \ent\left(\pi_i(\cdot \mid s_t)\right) \;\middle|\; s_0 = s \right],
\]
where the expectation is over game transitions and the joint policy, $\beta$ is the entropy coefficient, $\ent(\pi) = -\sum_a \pi(a)\log \pi(a)$ the entropy of $\pi$. The corresponding Q-function is
\[
    Q^\bpi_{i}(s, \ba) \coloneqq R_i(s,a_i) + \gamma \sum_{s'} \transfunc(s' \mid s, \ba ) V^\bpi_{i}(s').
\]
We further denote $\bar Q^\bpi_i(s,a) = \sum_{\ba_{-i}} \bpi_{-i}(\ba_{-i}\mid s) Q^\bpi_i(s, \ba)$ the expected Q-value over the other agents' policies. 

\paragraph{Agents and groups.}
An agent $i$ is an individual decision-maker with its own reward function $R_i$, and an entropy parameter $\beta_i$ controlling its stochasticity. For simplicity, we assume all agents share the same entropy parameter $\beta_i = \beta$. 
A group $\group$ is a subset of agents of fixed size $n$, i.e. $\group \subseteq \{1, \dots, m\}$, where $m$ is the number of available agents.
For a group $\group$ playing a game, we define its joint policy $\bpi_\group = (\pi_{\group,i})_{i \in \group}$ and joint reward $\bR_\group = (R_{\group,i})_{i \in \group}$, where the group subscript on $\pi_{\group,i}$ and $R_{\group,i}$ indicates both policies adopted and effective rewards are group dependent.

\paragraph{Optimality.}
In single-agent settings, a policy is commonly defined optimal if it maximises expected cumulative rewards. In multi-agent games, however, this notion becomes insufficient, since each agent's objective depends on the strategies of others. A more meaningful solution concept is the Quantal Response Equilibrium (QRE), an entropy-regularised equivalent of the Nash Equilibrium. Formally, agents play a QRE $\bpi^*$ if no agent can unilaterally improve their regularised value by deviating, assuming the other agents' policies remain fixed. That is, for every state $s$:
\begin{equation}
    V^{\bpi^*}_i(s) \geq V_i^{\{\pi_i\} \cup \bpi^*_{-i}}(s), \quad \forall \pi_i \in \Pi.
\end{equation}
Suboptimality can manifest in two ways: (1) as the \emph{distance from a QRE}, reflecting that some agents could improve their regularised value by adjusting their policies, and (2) through \emph{higher stochasticity}~\citep{laidlaw2022boltzmann, chan2021scalable}. QRE policies are optimal under entropy-regularisation, while in the limit $\beta \to \infty$, they approach to standard Nash equilibrium, corresponding to `raw' optimality.

\section{Problem Formulation}
\label{sec:setting}

We study a group $\group$ of $n$ agents interacting in a Markov game $\mg$.  
We are given a set of demonstrations
\[
    \demos_\group = \{\tau_k\}_k, 
    \quad \tau_k = (s_0^k, \mathbf{a}_0^k, s_1^k, \mathbf{a}_1^k, \dots),
\]
where each trajectory $\tau_k$ is generated by the agents following the joint policy $\bpi_\group$.  
Each agent $i \in \group$ is assumed to act near-optimally with respect to their unknown, individual reward function $R_i$.

\subsection{Altruism-Structured Rewards}

To capture the social nature of agent interactions, we assume that each agent's reward depends not only on its own intrinsic preferences but also on the outcomes experienced by other agents~\citep{sawyer1966altruism}.  

\begin{assumption}[Altruism-structured rewards]
\label{ass:altruism}
Each agent $i$ has an \emph{intrinsic reward function} $r_i : \states \times \actions^n \rightarrow [r_\text{min}, r_\text{max}]$, and an \emph{altruism level} $\lambda_i \in [\lambda_{\min}, \lambda_{\max}]$.
The effective reward of agent $i$ when acting within group $\group$ is
\[
    R_{\group, i}(s, \mathbf{a}) \;=\; r_i(s, a_i) \;+\; \frac{\lambda_i}{n-1} \sum_{\substack{k \in \group \\ k \neq i}} r_k(s, a_k).
\]
\end{assumption}

\begin{figure}[t]
\centering
\begin{tikzpicture}[scale=0.35, line cap=round]
  \node[anchor=center] at (0, 2.2) {{Altruism level $\lambda$}};

  \draw[very thick] (-10,0) -- (-9,0);
  \draw[dotted, very thick] (-7,0) -- (-9,0);
  \draw[very thick] (-7,0) -- (7,0);
  \draw[dotted, very thick] (7,0) -- (9,0);
  \draw[very thick,->] (9,0) -- (10,0);

  \node[left]  at (-10,0) {$-\infty$};
  \node[right] at (10,0) {$+\infty$};

  \foreach \x/\val/\label in {
    -5/--1/Antisocial,
     0/0/Egoistic,
     5/1/Prosocial
  }{
    \draw[thick] (\x,0.15) -- (\x,-0.15);
    \node[above=6pt] at (\x,0) {\val};
    \node[below=6pt] at (\x,0) {\small \label};
  }

  \node[below=6pt] at (-9.5,0) {\small Adversarial};
  \node[below=6pt] at (9.5,0) {\small Altruistic};

\end{tikzpicture}
\caption{The altruism scale model. We highlight three key values of $\lambda$. $\;\mathbf{-1}\;$: agent values its own welfare as much as it harms others. $\;\mathbf{0}\;$: agent ignores others' welfare. $\;\mathbf{1}\;$: agent values its own and others' welfare equally. \label{fig:continuum}}

\end{figure}
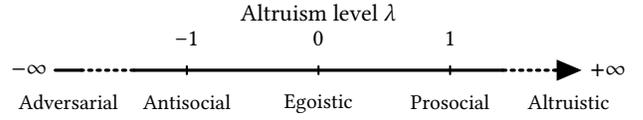

This simple yet expressive formulation captures the trade-off between self-interest and concern for others. Positive values of $\lambda_i$ correspond to altruistic behaviour, $\lambda_i = 0$ to purely selfish agents, and negative values reflect adversarial tendencies. Figure~\ref{fig:continuum} illustrates the continuum of behaviours induced by different levels of altruism. The model makes the following simplifying assumptions on social interactions:
\begin{enumerate}[topsep=4pt, leftmargin=12pt]
    \item Agents have access to each other's intrinsic rewards, which is reasonable whenever familiar individuals interact. This assumption is important, because otherwise agent behaviour would change as they learned more about each other.
    \item An agent's altruism level is independent of who it interacts with. This allows us to capture social tendencies. For example, individuals can be globally pro-social.  While this assumption can be relaxed, it allows us to keep the setting simple.
   \item Each agent's intrinsic reward only depends on that agent's action and the state. This is not a restrictive assumption. While rewards could depend on joint actions, this would not result in a richer setting, as any interacting terms can be modelled as part of the joint state.
\end{enumerate}

While in general, altruism-structured rewards induce general-sum games, we observe that in the specific cases where altruism levels are either all equal to $-1$ or $n-1$, we obtain zero-sum and fully cooperative game dynamics, respectively. Denoting $\mathbf{z}$ and $\mathbf{c}$ groups of antisocial and cooperative agents respectively, we have :
\[
    \sum_{i\in \mathbf{z}} R_{\mathbf{z},i}(s, \ba) = 0, \;\;\;\text{and}\;\;\; R_{\mathbf{c}, 1}(s, \ba) = \dots = R_{\mathbf{c}, n}(s, \ba).
\]
This insight indicates that altruism generalises from structured games to a wide variety of general sum games. This implies that our results cover a wide range of type of games.

\subsection{Objective}

Our goal is to interpret observed social behaviour. We therefore wish to recover each agent's ground-truth reward function $R_i$ from the demonstrations $\demos_\group$, under the assumption that these rewards are altruism-structured.  
Formally, we aim to disentangle each agent's reward into its intrinsic component and altruism level $R_i = (r_i, \lambda_i)$. By recovering these components, we obtain a concise and interpretable description of each agent's preferences and social tendencies, which generalises across different games and interaction contexts.


\section{Identifying Altruism and Intrinsic Rewards}
\label{sec:identifiability}

Before diving into reward inference, we must ask: when agents are altruistic, can we disentangle intrinsic rewards from social incentives? Altruism entangles agents' motivations, possibly complicating inference. We show that even when altruism levels are known, reward ambiguity persists unless agents are observed in multiple interaction groups. Interestingly, hiding altruism adds no further ambiguity under the same conditions.

\subsection{QRE policies}
\label{subsec:uniqueness}

To reason formally about reward identifiability, we first examine the structure of QRE policies under entropy-regularised reinforcement learning. As noted in Section~\ref{sec:prelims}, entropy-regularisation admits QREs as equilibria, which, unlike general-sum Nash equilibria, are unique for a given reward function. This uniqueness simplifies the problem of tracking feasible rewards, as it removes ambiguity over which equilibrium is observed~\citep{freihaut2025feasiblerewardsmultiagentinverse}. Formally, QRE policies have the following characterisation:
\begin{equation}
\label{eq:qre}
\pi^*_i(s,a) = \frac{\exp\left(\beta \bar Q_i^{*}(s, a)\right)}{\sum_{a'} \exp\left(\beta \bar Q^*_i(s, a')\right)}.
\end{equation}
Note that the setting introduces entropy parameters into the inference process. For the purpose of the analysis presented in this section, we will assume it is known. We later show in Section~\ref{sec:inference} that this poses no fundamental obstacle: we can infer over them using a suitable prior of the parameter. Later, we will remove the QRE assumption completely.

\subsection{Known Altruism}

We first consider the case where altruism levels are known.

\begin{proposition}
    \label{th:reward_ambiguity}
    Assume we observe a QRE $\bpi^*$ for the game $\mg(\bR)$, and that we know the altruism levels of agents.
    Then, intrinsic rewards are identifiable up to potential shaping transformations $\tilde r_i(s,a) = r_i(s,a) + \delta{r_i}(s,a)$, with
    \[
        \delta{r_i}(s,a_i) = \gamma \sum_{s'}\transfunc^{\bpi^*_{-i}}_{a_i}(s')\phi(s') - \phi(s),
    \]
    where $\phi: \states \rightarrow \mathbb{R}$ is any potential shaping function.
\end{proposition}
Even with known altruism, intrinsic rewards remain difficult to identify, mirroring the same ambiguities encountered in single-agent IRL~\citep{ng1999theory}. To reduce the remaining ambiguity, one approach in single-agent IRL is to observe multiple environments~\citep{cao2021identifiability}. In our multi-agent setting, we recycle the idea by constructing different transition dynamics through \emph{agent groups}. For example, in a 2-player game with 3 agents, we can observe up to three distinct equilibria instead of just one. These multiple equilibria can then be exploited for identification: if a candidate reward explains one group's behaviour but not another's, it can be ruled out. This idea, observing agents across diverse groups to constrain feasible rewards, is novel in MAIRL. Specifically, by observing an agent in two groups that induce sufficiently different transition dynamics, we can reduce its reward ambiguity up to state specific shifts.

\begin{corollary}
    \label{th:reduced_ambiguity}
    Let $\group$ and $\group'$ be two distinct groups containing agent $i$, and assume we observe QREs $\bpi^*_\group$ and $\bpi^*_{\group'}$ with known altruism.
    Then, the intrinsic rewards of agent $i$, can be recovered up to some non-trivial state-dependent shifts $\tilde r_i(s,a) = r_i(s,a) + \delta{r_i}(s)$, 
    if and only if the rank condition
    \[
        \operatorname{rank}\; \!
                \left(\begin{matrix}
                I - \gamma \transfunc^{\bpi^*_{\group-i}}_{a^1} & I - \gamma \transfunc^{\bpi^*_{\group'-i}}_{a^1} \\
                \vdots & \vdots \\
                I - \gamma \transfunc^{\bpi^*_{\group-i}}_{a^{|\actions|}} & I - \transfunc^{\bpi^*_{\group'-i}}_{a^{|\actions|}}
                \end{matrix}\right)
            = 2|\mathcal{S}| - 1
    \]
    holds. If $\lambda_i = 0$, the intrinsic rewards can be identified up to a constant.
\end{corollary}

Even with two induced dynamics, some ambiguity can remain when an agent is not purely egoistic ($\lambda_i \neq 0$), because its consideration for others' intrinsic rewards introduces additional degrees of freedom. Intuitively, the intrinsic rewards of other agents must also be pinned down to fully resolve the ambiguity.

\begin{corollary}
\label{th:reward_recovery}
Under the conditions of Corollary~\ref{th:reduced_ambiguity}, if each agent is observed in two groups and the rank condition holds for every pair, the intrinsic rewards can be recovered up to a constant.
\end{corollary}

Importantly, forming multiple groups to meet Corollary~\ref{th:reward_recovery}'s requirement is surprisingly not expensive: introducing just one extra agent allows construction of groups of size $n$ such that, in the best case, only three distinct groups are needed to satisfy the rank condition for all pairs (three groups are enough to ensure that all agents played twice). With this, intrinsic rewards can be reliably recovered, allowing us to proceed to our main theoretical result on disentangling \emph{latent} altruism.

\subsection{Latent Altruism}

We now tackle the case where altruism is latent. Intuition suggests that unknown altruism makes inference much harder, but remarkably, simply observing agents across sufficiently diverse groups is enough to fully disentangle altruism from intrinsic rewards.

\begin{theorem}
    \label{th:full_recovery}
    Let us have a set of $n+1$ agents, and assume we observe enough QREs such that we observed each agent act in two separate groups $(\group_i, \group_i')$.
    Then, if and only if the rank condition from Corollary~\ref{th:reduced_ambiguity} is verified on every pair $(\group_i, \group_i')$, 
    the altruism levels can be perfectly disentangled from intrinsic rewards for all agents. Specifically, the altruism levels can be exactly identified, and the intrinsic rewards can be recovered up to a constant.
\end{theorem}

\paragraph{Intuition.} To see why Theorem~\ref{th:full_recovery} holds, consider perturbing agent $i$'s altruism: $\tilde \lambda_i = \lambda_i + \delta \lambda_i$. In principle, this change could be compensated by adjusting $i$'s intrinsic rewards, the intrinsic rewards of other agents, or the value function. However, the rank condition ensures that the value function can only absorb \emph{constant shifts} across states, while the effect of $\delta \lambda_i$ depends on the other agents' rewards in a state-dependent way, so it cannot be canceled by the value function.

Next, observing agent $i$ in multiple groups with different compositions rules out compensating via the intrinsic rewards of others: the perturbation interacts differently in each group. Likewise, agent $i$'s intrinsic reward cannot simultaneously offset $\delta \lambda_i$ across groups. Together, these constraints uniquely pin down $\lambda_i$, allowing us to disentangle altruism from intrinsic rewards before recovering the latter with Corollary~\ref{th:reward_recovery}.

\smallskip
While richer characterisations of transition dynamics exist~\citep{schlaginhaufen2024towards}, verifying the rank condition in practice is challenging, especially with sub-optimal demonstrations or unknown $\beta$, and we cannot freely choose agents to satisfy it. As multiple groups generally reduce ambiguity, we suggest adopting a more practical approach using Bayesian inference on groups of the agents available.

\section{Two Bayesian methods for inferring altruism and intrinsic rewards.}
\label{sec:inference}

We now turn to the practical question of how to infer intrinsic rewards and altruism from observed behaviour.
Crucially, we introduce methods that can ingest demonstrations collected from \emph{multiple groups}, leveraging results from Section~\ref{sec:identifiability}.

We adopt a \emph{Bayesian framework}, which offers a principled way to represent uncertainty over rewards~\citep{ramachandran2007bayesian}.
Given a prior $\bel(\cdot)$ over reward functions, Bayesian MAIRL/IRL seeks the posterior $\bel(\cdot \mid \demos)$ conditioned on demonstrations $\demos$.
We perform posterior sampling using SGLD~\citep{welling2011bayesian}, a gradient-informed sampler well-suited for efficiently exploring complex posteriors.\footnote{Further discussion of our motivation for this choice is provided in Appendix~\ref{appendix:algorithms}.}
We propose two complementary Bayesian approaches: (1) extending standard Bayesian IRL~\citep{jeon2020reward, buning2022interactive, buening2024environment} to the multi-agent setting with altruistic rewards, by evaluating the likelihood of observed demonstrations under QRE policies. (2) Our main contribution, which (similarly to~\citep{dimitrakakis2011bayesian}) first infers a posterior over policies given demonstrations, and then calculates reward posterior by marginalising over the inferred policies.

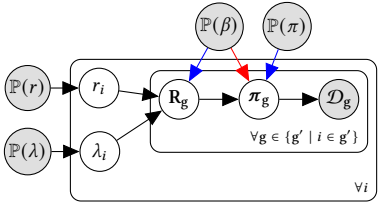
\begin{figure}
\centering
\begin{tikzpicture}[scale=0.75, every node/.append style={transform shape}]

  \node[obs]                              (pr) {$\bel(r)$};
  \node[obs, below=0.3cm of pr]          (pl) {$\bel(\lambda)$};

  \node[latent, right=0.5cm of pr]            (ri) {$r_i$};
  \node[latent, right=0.5cm of pl]          (li) {$\lambda_i$};

  \node[latent, right=0.7cm of ri, yshift=-0.2cm]            (Rg) {$\bR_\group$};
  \node[latent, right=0.7cm of Rg]          (Pg) {$\bpi_\group$};

  \node[obs, right=0.7cm of Pg]             (demo) {$\demos_\group$};

  \node[obs, above=0.5cm of Pg, xshift=0.5cm]          (pp) {$\bel(\pi)$};
  \node[obs, above=0.5cm of Pg, xshift=-0.7cm]         (pb) {$\bel(\beta)$};

  \plate {groups} {(Rg)(Pg)(demo)} {$\forall \group \in \{\group' \mid i \in \group' \}$};
  \plate {agents} {(ri)(li)(groups.north west)(groups.south east)} {$\forall i$};

  \edge {pr} {ri};              
  \edge {pl} {li};                

  \edge[color=blue] {pp} {Pg};     
  \edge[color=red] {pb} {Pg};  
  \edge[color=blue] {pb} {Rg};                                            
  \edge {ri, li} {Rg};            
  \edge {Rg} {Pg};        
  \edge {Pg} {demo};

\end{tikzpicture}

\caption{Graphical model of policy and altruism-structured rewards across groups. Latent nodes are white. Priors $\bel(r)$ and $\bel(\lambda)$ generate agent rewards $r_i$ and altruism $\lambda_i$, which determine group rewards $\bR_\group$ and policies $\bpi_\group$, producing demonstrations $\demos_\group$. Priors $\bel(\pi)$ and $\bel(\beta)$ also participate in generating rewards and policies. 
Coloured edges indicate how and which prior is used (red is for DRP, blue for PORP). \label{fig:bayesnet}}

\Description{Bayesian network illustrating how demonstrations from multiple groups influence the rewards of individual agents.}
\end{figure}

In the following, we consider demonstrations from multiple groups. We denote $\demosall = \bigcup_\group \demos_\group$ the full demonstration set, and $\Rset = \left(R_i\right)_{i=1}^m = \left((r_i, \lambda_i)\right)_{i=1}^m$ the unknown reward vector where $m$ is the total number of agents observed. Altruism is assumed independent of intrinsic rewards, and demonstrations are independent across groups. Figure~\ref{fig:bayesnet} illustrates the overall inference process.

\subsection{Direct Reward Posterior (DRP)}
\label{subsec:direct_reward_posterior}


One way to write the reward posterior over rewards is to directly model the likelihood of the demonstrations under those rewards, i.e., $\bel(\demosall \mid \Rset) = \prod_{\tau \in \demosall} \bel(\tau \mid \Rset)$. Since demonstrations are independent across groups, we can rearrange the product accordingly:
\[
\bel(\demosall \mid \Rset) = \prod_{\group}\prod_{\tau \in \demos_\group} \bel(\tau \mid \bR_\group).
\]

This formulation is conceptually similar to that of~\citet{buening2024environment}, though their focus is on multiple environments, whereas we consider different agent groups. Somehow, we need to specify the likelihood $\bel(\tau \mid \bR)$, modelling how agents generate trajectories based on rewards. 
We propose to model demonstrated policies as QRE. In single-agent settings, it reduces to the standard Boltzmann form used in IRL. We therefore assume all agents adopt QRE policies, and denote by $\bpi^*_{\bR}(\,\cdot \, ; \beta)$ the QRE policy under joint reward $\bR$ with entropy parameter/optimality $\beta$. Given a prior belief $\bel(\beta)$ over $\beta$, we can marginalise it out, and obtain the trajectory likelihood:
\begin{equation*}
    \bel(\tau \mid \bR) = \prod_{(s, \ba) \in \tau} \int_{\beta{_\text{min}}}^\infty \bpi^*_{\bR}( \ba \mid s; \beta) \text{d}\bel(\beta),
\end{equation*}
with $\beta_{\min} > 0$. Substituting this expression back into the demonstration likelihood, Bayes'rule yields the full DRP:
\begin{equation}
\label{eq:direct_posterior}
    \bel(\Rset \mid \demosall) \propto \bel(\Rset) \prod_{\group} \prod_{\tau \in \demos_\group} \prod_{(s, \ba) \in \tau} \int_{\beta_\text{min}}^\infty \bpi^*_{\bR_\group} (\ba \mid s; \beta) \text{d}\bel(\beta).
\end{equation}
where the prior over rewards can be expressed as a product of priors over altruism and intrinsic rewards $\bel(\Rset) = \prod_i^m \bel(r_i)\bel(\lambda_i)$.
We next propose a novel reward posterior that bypasses the need to model policies as QREs.

\subsection{Policy-Oriented Reward Posterior (PORP)}
\label{subsec:porp}

DRP evaluates trajectory likelihoods assuming agents adopt QRE policies. It is both a strong assumption, and intractable as computing QREs grows expensive. In contrast, in this section we place a prior over policies, and infer a posterior over policies from the demonstrations. This allows us to simply use a prior on the \emph{suboptimality} of the policies demonstrated, rather than assume a behavioural model such as QRE.
Within any group, we can express the posterior over rewards by marginalising over joint policies:
\begin{equation}
    \label{eq:porp_init}
    \bel(\bR \mid \demos) = \int_{\Pi^n} \bel(\bR \mid \bpi) \text{d} \bel(\bpi \mid \demos),
\end{equation}
assuming that rewards are conditionally independent of demonstrations given policies $\bel(\bR \mid \bpi, \demos) = \bel(\bR \mid \bpi)$. This requires us to specify how likely a reward is, given a policy. Intuitively, a reward is more plausible if the policy is nearl-optimal under it. We formalise this via a \emph{gap function} $\Delta_{\bR, \beta} : \Pi \rightarrow \mathbb{R}^+$, which measures the suboptimality of policy $\bpi$ under joint reward $\bR$ for a specific entropy parameter $\beta$. The likelihood of $\bR$ given $\bpi$ and $\beta$ is then:
\begin{equation}
\label{eq:gap_posterior}
\bel(\bR \mid \bpi, \beta) = \frac{1}{Z_{\bpi, \beta}} \bel(\bR) \cdot e^{-c \Delta_{\bR, \beta}(\bpi)},
\end{equation}
where $Z_{\bpi, \beta}$ is a partition function, and where $c$ controls how strictly we believe agents are near-optimal. Plugging this into \eqref{eq:porp_init} and marginalising over $\beta$, we obtain:
\begin{equation*}
    \bel(\bR \mid \demos) =  \int_{\Pi^n} \frac{1}{Z_{\bpi}} \int_{\beta{_\text{min}}}^\infty \bel(\bR) \cdot e^{-c \Delta_{\bR, \beta}(\bpi)} \text{d} \bel(\beta)\text{d} \bel(\bpi \mid \demos),
\end{equation*}
Essentially, this posterior says a reward is more plausible if the observed demonstrations could have come from policies that are nearly optimal under it, averaged over all plausible policies and levels of stochasticity across groups.
A practical difficulty here is that $Z_\bpi$ varies with each policy, and thus cannot be pulled outside of the integral over policies. However, we show that using an appropriate gap function, we get $Z_\bpi \approx Z$ across policies, partly because the posterior over policies $\bel(\bpi \mid \demos)$ tends to be sharply peaked, especially when enough demonstration data is available.\footnote{Both a theoretical and empirical study on the near constant nature of $Z_\bpi$ is provided in Appendix~\ref{apx:results}.} This makes the following approximation reasonable:
\begin{equation}
    \bel(\Rset \mid \demosall) \appropto \prod_\group \int_{\Pi^n}\int_{\beta{_\text{min}}}^\infty \bel(\bR_\group) \cdot e^{-c \Delta_{\bR_\group, \beta}(\bpi)} \text{d} \bel(\beta)\text{d} \bel(\bpi \mid \demos_\group).
    \label{eq:porp}
\end{equation}
This posterior is easy to compute, as we show in the next paragraph. We then only need to choose an appropriate gap function, as we explain in the remainder of this section.
\paragraph{Two-step posterior sampling.}
We now outline a practical procedure to sample from the PORP \eqref{eq:porp}. An SGLD-based implementation is provided in Appendix~\ref{appendix:algorithms}. The method proceeds in two steps:
\begin{enumerate}[topsep=4pt, leftmargin=12pt]
    \item  
    Given that the likelihood of a trajectory under a policy is simply the probability of the policy generating it, we have: $\bel(\bpi \mid \tau) \propto \bel(\bpi) \prod_{(s,\ba) \in \tau} \bpi(\ba \mid s)$. Factoring over a group's dataset:
    \begin{equation}
    \bel(\bpi \mid \demos) \propto \prod_{\tau \in \demos} \bel(\bpi) \prod_{(s, \ba) \in \tau} \bpi(\ba \mid s).
    \label{eq:policy_posterior}
    \end{equation}
    In the first step, we draw $N$ samples ${\hat \bpi_{\group, 1}, \dots, \hat\bpi_{\group, N}}$ from this posterior for each group $\group$.
    
    \item We estimate the reward posterior \eqref{eq:porp} using Monte Carlo integration:
    \begin{equation}
        \bel(\Rset \mid \demosall) \appropto \prod_\group \sum_{k=1}^N \int_{\beta{_\text{min}}}^\infty \bel(\bR_\group) \cdot e^{-c \Delta_{\bR_\group, \beta}(\hat\bpi_{\group, k})} \text{d} \bel(\beta).
        \label{eq:mc_porp}
    \end{equation}
    We are then left with sampling from that approximate posterior to generate reward candidates.
\end{enumerate}
This two-step approach enables efficient approximation of the posterior over rewards: first by inferring plausible policies from data, and then by weighting rewards according to how well these policies align with optimal behaviour under each reward hypothesis.

\paragraph{Gap functions.}
Gap functions are constructed such that a value of zero indicates an equilibrium, while larger values reflect deviations from optimality. The Nash Imitation Pap~\citep{ramponi2023imitation}, recently standardised in the MAIRL setting~\citep{freihaut2025feasiblerewardsmultiagentinverse}, satisfies this property. It measures the maximum incentive any agent has to unilaterally deviate from the current policy. A gap of zero implies that no agent can benefit from deviating, thus identifying a Nash equilibrium. We extend this to the entropy-regularised setting.
\begin{definition}[QRE Imitation Gap (QIG)]
    The QIG is defined as:
    \begin{equation}
        \gapv_{\bR, \beta}(\bpi) \coloneqq \max_{i\in[n]}\max_{\pi_i \in \Pi}\sum_{s\in\states}V_i^{\{\pi_i\} \cup \bpi_{-i}}(s) - V_i^{\bpi}(s),
    \end{equation}
    where the values are computed on $\bR$ and $\beta$.
\end{definition}
Compared to DRP, the QIG requires computing best responses instead of full equilibria. However, as the number of players grows, even the QIG can become computationally prohibitive. To address this, we introduce the Policy Stability Gap (PSG), which exploits the known structure of optimal entropy-regularised policies.
\begin{definition}[Policy Stability Gap (PSG)]
    Let $\dkl$ be the Kullback-Leibler (KL) divergence. Furthermore, let the `soft response' of the $i$-th agent to the joint policy $\bpi_{-i}$, under rewards $\bR$ and regularisation parameter $\beta$, be:
    \begin{equation*}
        \sigma_\bR^{\bpi_{-i}}(a \mid s; \beta) \coloneqq \frac{\exp{\left(\beta \bar Q^\bpi_i(s, a)\right)}}{\sum_{a'}\beta \exp{\left(\beta \bar Q^\bpi_i(s, a')\right)}}
    \end{equation*}
    where the Q-values are computed on $\bR$ and $\beta$. We define the PSG as:
    \begin{equation}
            \gapkl_{\bR, \beta}(\bpi) \coloneqq \max_{i\in[n]} \sum_{s\in\states}\dkl \Big( \pi_i(\cdot \mid s) \;\bigl\vert\bigl\vert\; \sigma_\bR^{\bpi_{-i}} (\cdot \mid s; \beta ) \Big).
    \end{equation}
\end{definition}

The PSG measures how far a joint policy deviates from its soft response. A QRE policy inherently has a PSG of 0, since it is equal to its soft response by definition (Eq.\eqref{eq:qre}).
Crucially, PSG is computationally efficient, requiring only the evaluation of $Q$ for the current joint policy.

It is unclear whether these gap functions truly capture human irrationality. Nonetheless, they naturally capture suboptimality: small gaps indicate that agents act largely rationally but occasionally make mistakes. By measuring deviations from best (QIG) or soft (PSG) responses, we can infer rewards from demonstrations \emph{without} assuming perfect QRE behaviour.

\subsection{Priors}

To enable proper and stable Bayesian inference, we need to specify the priors. We place independent Gaussian priors on intrinsic rewards and policies. The reward prior is centered at zero, reflecting the assumption that states with little information carry no inherent reward, an assumption that is generally safer than positing there are incentives in unreachable states. Similarly, the policy prior is centered around uniform action probabilities, capturing the idea that, in the absence of evidence, agents behave without clear preference.
For altruism, we use a uniform prior to enable exploration across the entire altruism spectrum, though more informative choices (e.g., a Gaussian centered around egoism) could be employed when empirical data are available.
Finally, we assign an exponential prior to the entropy parameter $\beta$, assuming demonstrations are mostly low-entropy, reflecting confident yet occasionally inconsistent behaviour. Further details are provided in Appendix~\ref{apx:details}.


\section{Experiments}
\label{sec:experiments}

Our experiments aim to answer two key questions: (1) Can we \emph{disentangle altruism from intrinsic rewards} by observing agents in different groups? (2) Can we synthesise behaviours corresponding to \emph{any level of altruism} by doing so ?

To address the first question, Section~\ref{sec:random_mgs} studies challenging randomised MGs, providing a robust benchmark and enabling ablation studies on using multiple groups of agents for demonstrations.
To answer the second question, Section~\ref{sec:collaborative_cooking} considers a scenario where kitchen employees with conflicting tendencies act together, testing whether we can generate altruistic behaviours from inherently anti-social agents.
We compare our approach against state-of-the-art maximum entropy and inverse Q-learning methods, MAAIRL~\citep{song2018multi} and MAMQL~\citep{haynam2025multi}, which we adapt to use an altruism-structured reward model. These baselines, however, are not extended to multiple-group inference.
To ensure a fair comparison, all methods are provided with the same total number of demonstrations. For methods using multiple groups, the demonstration budget is evenly split across all available groups (algorithms do not choose which group they use). Trajectories have a fixed length of 1000 timesteps, and all metrics are computed with respect to the first group, which is shared across methods. Demonstrations are generated from QRE policies under unknown stochasticity $\beta$. For MAMQL and MAAIRL, which do not explicitly account for $\beta$, we provide the true value.
For all of our experiments, we set $\lambda_{\text{max}} = -\lambda_{\text{min}} = 5$, $r_{\text{max}} = 1$ and $r_{\text{min}} = 0$.
All errors are reported as mean squared errors, rescaled such that $1$ corresponds to the expected error of random guessing, with standard errors included.
Note that our experiments do not attempt to address MAIRL scalability in general. Instead, we focus on the largest environments that can be handled without approximations such as neural networks, leaving broader scalability questions to future work. Additional results and experimental details, including priors, are provided in Appendices~\ref{apx:results} and \ref{apx:details}.

\subsection{Disentangling Altruism from Intrinsic Rewards}
\label{sec:random_mgs}

In this experiment, we validate DRP and PORP on sets of non-trivial randomised MGs. 
\paragraph{Experimental setup.} 
We consider two sets of randomised MGs:  
\begin{enumerate}[topsep=4pt, leftmargin=12pt]
    \item A 3-player, 512-state, 5-action set, which serves as our main benchmark domain. Here, we provide 4 agents, allowing methods to infer from 4 different combinations of agents. We provide algorithms with a budget of 200 trajectories.
    \item A 4-player, 16-state, 5-action set, with a pool of 6 agents, to study the benefits of using a subset or all possible groups. We also evaluate methods on an increasingly large budget for demonstrations.
\end{enumerate}
Transition functions are sampled from a Dirichlet distribution. 
Each agent's altruism parameter is sampled uniformly, while intrinsic rewards are defined by randomly selecting a subset of state-action pairs with reward 1, with all other pairs set to 0. Those are naturally hidden from algorithms. Metrics are averaged over 10 seeds.

\paragraph{Results.}

\begin{figure}[t]
    \centering
    \includegraphics[width=0.99\linewidth]{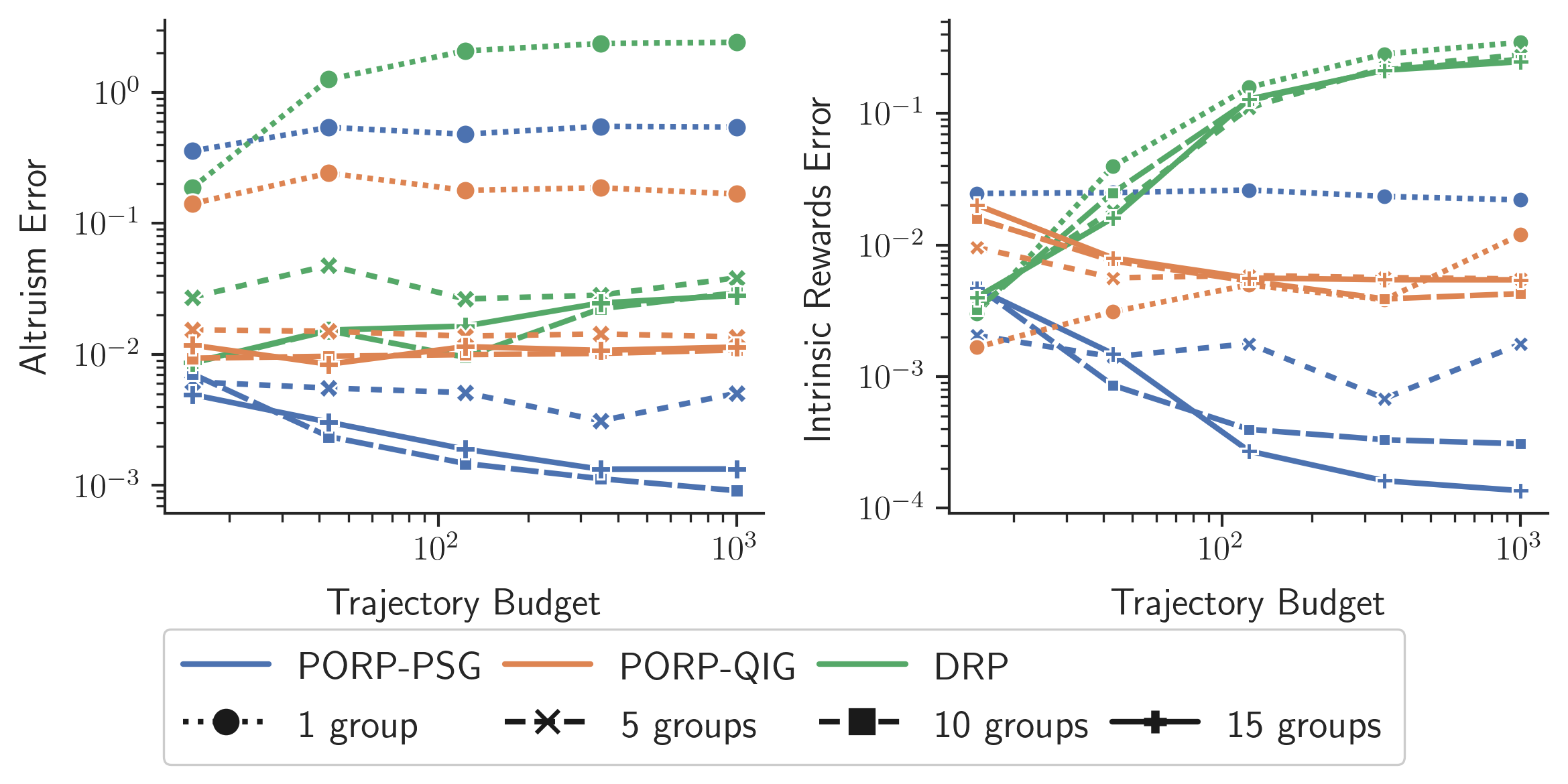}
    \caption{Sample efficiency of the proposed Bayesian inference methods on 4-player randomised games, using different numbers of groups. Error bars are provided in Appendix~\ref{apx:results}.}
    \label{fig:random_mg_sample}
\end{figure}

Table~\ref{tab:benchmark_results} and Figure~\ref{fig:random_mg_sample} both demonstrate that PORP-PSG is the most reliable method for disentangling altruism from intrinsic rewards, consistently achieving much lower errors than all other approaches. Figure~\ref{fig:random_mg_sample} further shows that inferring altruism and intrinsic rewards is extremely difficult without multiple groups. Introducing as few as 5 groups substantially reduces ambiguity. When increasing the number of groups, there appears to be a trade-off between the information gained from additional groups and the information lost due to fewer demonstrations per group.
DRP performs surprisingly well in 3-player games using a single group (Table~\ref{tab:benchmark_results}) but exhibits limited robustness in 4-player environments (Figure~\ref{fig:random_mg_sample}). Its performance declines as trajectory budget increases, likely because fitting QREs becomes harder with larger datasets. Finally, using the QIG gap function improve over basic baselines but remain far behind PSG.  
Overall, PORP with PSG emerges as the most promising method, combining strong performance, high sample efficiency, and low computational cost. Additional results on randomised MGs are provided in Appendix~\ref{apx:results}.

\begin{table}[t]
\centering
\caption{Altruism and intrinsic reward recovery errors on Random MGs. Lower is better.}
\small
\vspace{0.5em}
\begin{tabular}{lcc}
\toprule
\textbf{Method} & \textbf{Altruism Error} & \textbf{Intrinsic Rewards Error} ($\times 10^3$) \\
\midrule
PORP-PSG & \textbf{0.024 $\pm$ 0.003} & \textbf{0.065 $\pm$ 0.004}\\
PORP-QIG & 0.392 $\pm$ 0.156 & 0.338 $\pm$ 0.081\\
DRP & 0.040 $\pm$ 0.007 & 0.091 $\pm$ 0.005 \\
\hdashline
MAMQL & 0.855 $\pm$ 0.156 & 1.526 $\pm$ 0.041 \\
MAAIRL & 3.313 $\pm$ 0.435 & 1.555 $\pm$ 0.042 \\
\midrule
\multicolumn{3}{l}{\textit{Ablations (without groups)}} \\
\midrule
PORP-PSG & 0.352 $\pm$ 0.133 & 0.474 $\pm$ 0.134\\
PORP-QIG & 1.431 $\pm$ 0.118 & 1.003 $\pm$ 0.062 \\
DRP & 0.059 $\pm$ 0.008 & 0.120 $\pm$ 0.007 \\
\bottomrule

\end{tabular}
\label{tab:benchmark_results}
\vspace{-0.5em}
\end{table}

\begin{figure*}[t]
    \centering
    \begin{subfigure}[b]{0.75\linewidth}
        \centering
        \includegraphics[width=0.95\linewidth]{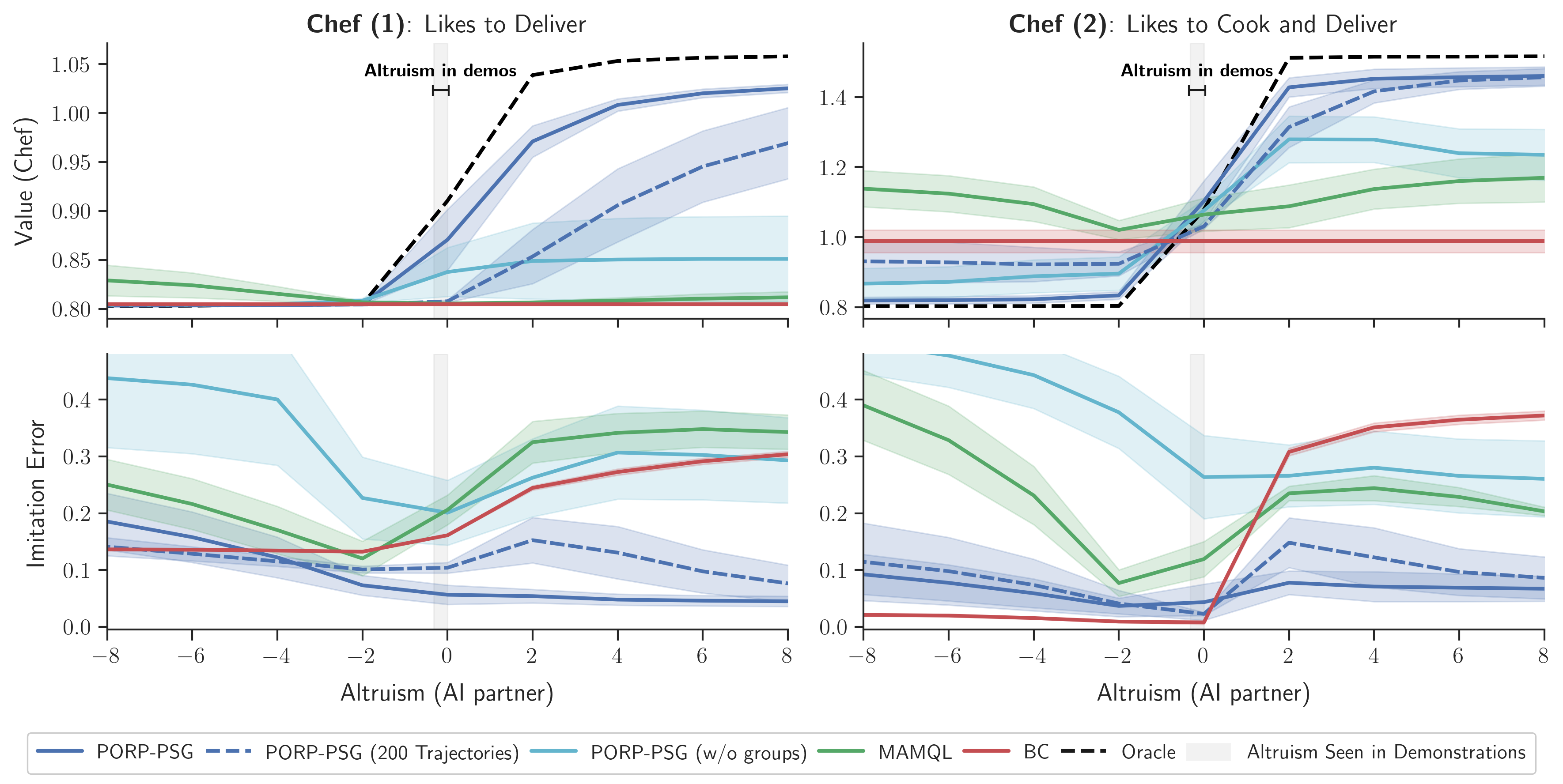}
        \caption{Imitation error of synthesised policies for various altruism levels and the value of their partner.}
        \label{fig:behaviour_plot}
    \end{subfigure}%
    \hfill
    \begin{subfigure}[b]{0.19\linewidth}
        \centering
        \includegraphics[width=\linewidth]{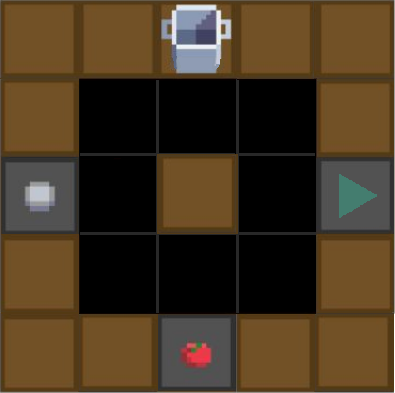}
        \vspace{1.65cm}
        \caption{Kitchen layout.}
        \label{fig:layout}
    \end{subfigure}
    \caption{Synthesis of policies from anti-social demonstrations in Overcooked, ranging from adversarial to altruistic.}
    \label{fig:behaviour_synthesis}
\end{figure*}

\subsection{Achieving Generalised Altruism Imitation in a Collaborative Cooking Task}
\label{sec:collaborative_cooking}

We consider a simplified kitchen scenario inspired by Overcooked~\citep{carroll_utility_learning_about_2019}, with three chefs, of which only two work per shift. Interpersonal tensions cause some chefs to act unreasonably, reducing productivity. Our goal is to improve welfare by identifying each chef's intrinsic rewards and introducing an AI partner that can exhibit any desired level of altruism toward its teammate.

\paragraph{Experimental setup.}
We design a compact but challenging kitchen (Figure~\ref{fig:layout}) with 3420 states and 5 actions (4 for moving, and 1 for interacting). Agents can cooperate (e.g., passing ingredients) or act adversarially (e.g., blocking the cooking pot). Each chef have their own intrinsic rewards: chef (1) receives $+1$ for \emph{delivering}, chef (2) for both \emph{cooking and delivering}, and chef (3) for \emph{cooking}. To assess how well the inferred rewards generalise to novel altruism levels, chefs latent altruism parameters are exclusively sampled uniformly from $[-0.25, 0]$ (slightly antisocial profiles).
Each algorithm receives 1000 trajectories, allocated either to a single group (1,2) or across all three groups (1,2), (2,3), and (1,3), from which intrinsic rewards $\hat r_i$ are inferred. We reconstruct joint reward functions by plugging a tunable altruism level $\lambda^{(\text{target})}$. We then replace either chef 1 or 2 in the (1,2) group with an AI partner optimising $ \hat R_{(1,2),1}(s,\ba) = \hat r_1(s,a_1) + \lambda^{(\text{target})} \hat r_2(s,a_2)$ or $\hat R_{(1,2),2}$ analogously, and let the remaining chef optimise their own true intrinsic reward egoistically. Varying $\lambda^{(\text{target})}$ allows the AI to imitate any altruism level towards the chefs.  
We measure the imitation quality through the chef's policy \emph{value} when partnered with the AI, compared to an oracle agent using ground-truth rewards, and the \emph{policy imitation error}, measured as the mean KL divergence from the oracle policy.
Due to the prohibitive computation requirements of DRP and QIG, we do not include them in this study. Instead, we include a Behaviour Cloning (BC) baseline, which replicates demonstration policies while allowing the partner to best respond. Experiments are repeated across five random seeds.

\begin{table}[t]
\centering
\small
\setlength{\tabcolsep}{2pt}
\caption{Reward and policy recovery errors on Overcooked. The chef value error measures the difference in the chef's policy value when paired with the AI versus the oracle, averaged over the full altruism spectrum. Lower is better.}
\vspace{0.5em}
\begin{tabular}{lcccc}
\toprule
\textbf{Method} & \textbf{Altruism} & \textbf{Intrinsic} & \textbf{Policy} & \textbf{Chef} \\
& \textbf{Err.} & \textbf{Rewards Err.} & \textbf{Imitation Err.} & \textbf{Value Err.} \\
\midrule
PORP-PSG & \textbf{0.008 $\pm$ 0.001} & \textbf{0.016 $\pm$ 0.001} & \textbf{0.082 $\pm$ 0.007} & \textbf{0.025 $\pm$ 0.005} \\
PSG \tiny{(w/o groups)} & 0.076 $\pm$ 0.004 & 0.225 $\pm$ 0.014 & 0.340 $\pm$ 0.018 & 0.330 $\pm$ 0.050 \\
\hdashline
MAMQL & 0.080 $\pm$ 0.002 & 0.330 $\pm$ 0.005 & 0.254 $\pm$ 0.011 & 0.450 $\pm$ 0.050 \\
BC & -- & -- & 0.188 $\pm$ 0.013 & 0.548 $\pm$ 0.055 \\
\bottomrule
\end{tabular}
\label{tab:overcooked}
\vspace{-0.5em}
\end{table}

\paragraph{Results.}
Table~\ref{tab:overcooked} shows that PORP-PSG can reliably imitate \emph{any desired level of altruism}, although it learned exclusively from antisocial demonstrations. 
Remarkably, as illustrated in Figure~\ref{fig:behaviour_plot}, the method hugs the oracle's curve, indicating near perfect imitation on the whole altruism continuum.
By contrast, MAMQL exhibits a striking misalignment: as the AI's intended altruism increases, the partner's utility does not improve, and in some cases even declines. This reveals a critical pitfall of incomplete reward disentanglement, where models tuned to `be kind' may in fact behave adversarially. 
Even under reduced data, PORP-PSG consistently produces agents whose altruism reliably translates into higher welfare for chefs, demonstrating both robustness and genuine social alignment.

\section{Conclusion}

We have demonstrated that observing agents across multiple groups is crucial for disentangling altruism from intrinsic rewards in MAIRL. Importantly, such multi-group demonstrations often arise naturally, incurring little extra cost. We propose a practical Bayesian method for constructing a posterior over policies from these demonstrations, we can infer accurate, disentangled rewards without restrictive assumptions on agent rationality. This enables \emph{generalised altruism imitation}, yielding socially aligned AI agents that are both more interpretable and trustworthy. Our results pave the way for AI systems that can \emph{reliably understand and replicate social behaviours}, providing a principled foundation for aligned, cooperative multi-agent interactions.

While our current model assumes context-independent altruism and perfect knowledge of others' preferences, these simplifications point to exciting future directions. Extensions include learning from learning agents, modelling altruism as a function of state and action histories, actively selecting informative groups, scaling to larger environments without dynamics model access~\citep{chan2021scalable}, and leveraging human proxies such as LLMs to model decision-making.



\bibliographystyle{ACM-Reference-Format} 
\bibliography{references}


\newpage
\onecolumn
\appendix

\begin{table}[h]
\centering
\caption{Notations}
\begin{tabular}{ll}
\toprule
\textbf{Symbol} & \textbf{Description} \\
\midrule

$\mg(\bR)$ & Markov game with joint rewards $\bR$\\
$m$ & Total number of agents \\
$n$ & Number of agents per game / group \\
$\group$ & Subset of agents forming a group, $|\group| = n$ \\
$a^i$ & Action index $i$ in the action space $\actions$ \\
$\beta$ & Entropy regularisation / suboptimality parameter \\
$\beta_\text{min}$ & Minimum $\beta$ considered when computing posteriors \\
$\Pi$ & Set of all policies \\
$\bpi_\group$ & Joint policy formed by the agents in group $\group$\\
$\bpi_{-i}$ & Joint policy $\bpi$ without agent $i$'s policy \\
$\pi_{\group, i}$ & Policy of the $i$-th agent in group $\group$\\
$r_{\group, i}$ & Intrinsic rewards of the $i$-th agent in group $\group$\\
$\hat \bpi$ & Inferred joint policy \\
$\hat r$ & Inferred intrinsic rewards \\
$\ent(\pi)$ & Policy Shannon Entropy \\
$R_{\group, i}$ & Reward of the $i$-th agent of a group $\group$\\
$\bar Q_i$ & Expected Q function for the $i$-th agent, w.r.t. the other agents policies\\
$\bar r_i$ & Expected intrinsic rewards generated by the $i$-th agent\\
$\Delta_{\bR, \beta}(\bpi)$ & Gap function measuring suboptimality of policy $\bpi$ under $\bR$ and $\beta$ \\
$c$ & Gap concentration parameter \\
$\demosall$ & Full set of demonstrations across all groups \\
$\demos_\group$ & Demonstrations corresponding to group $\group$ \\
$\bel(\cdot)$ & Priors \\
$\Rset_\psi$ & Full vector of reward parameters, parameterised by $\psi$ \\
$\bR_{\psi, \group}$ & Joint reward of agents in group $\group$, parameterised by $\psi$ \\
$\bpi_\theta$ & Joint softmax policy parameterised by $\theta$ \\
$\mathbf{P}_\group$ & Posterior samples of policies for group $\group$ \\
$\epsilon^t$ & Step sizes for SGLD updates \\
$T$ & Number of SGLD iterations\\
$w$ & Number of warm-up iterations for SGLD chains \\
$\eta_t$ & Gaussian noise in SGLD updates \\
\bottomrule
\end{tabular}
\label{tab:notation}
\end{table}

\newpage

\section{Omitted proofs}
\begin{proof}[\textbf{Proof of Proposition}~\ref{th:reward_ambiguity}]
    Observing the QRE $\bpi^*$ for the game $\mg(\bR_\group)$ implies that for each $i \in \group$,
    \[
    \pi^*_i(s,a) = \frac{\exp(\beta \bar Q^*_i(s,a))}{\sum_{a'\in\actions} \exp(\beta \bar Q^*_i(s,a'))},
    \]
    Specifically,
    \begin{align*}
        V^*_i(s) &= \sum_{a} \pi^*_i(s,a) \left[ \bar Q^*_i(s,a) - \frac{1}{\beta} \log \pi_i(a \mid s)\right]\\
                 &= \frac{1}{\beta} \log \sum_{a} \exp{\beta Q^*_i(s,a)}.
    \end{align*}
    Taking the log gives
    \begin{equation}
        \label{eq:logpi}
    \frac{\log \pi^*_i(s,a)}{\beta} = \bar Q^*_i(s,a) - V^*_i(s).
    \end{equation}
    We thus have for each agent
    \begin{equation}
    \label{eq:logpi_clean}
    \frac{\log \pi^*_i(s,a)}{\beta} = r_i(s,a) + \frac{\lambda_i}{n-1}\sum_{k\in\group\setminus\{i\}} \bar r_k(s) + \Delta V^*_i(s,a),
    \end{equation}
    where \(\Delta V^*_i(s,a) = \gamma \sum_{s'} \sum_{\ba_{-i}} \bpi_{-i}(\ba_{-i}\mid s)\transfunc(s'\mid s,\ba) V^*_i(s') - V^*_i(s)\), and $\bar r_{\group, k} = \sum_{a_k} \bpi^*_{\group, k}(a_k\mid s) r_k(s, a)$.

    \noindent Now, define perturbed rewards
    \[
    \tilde r_i(s,a) = r_i(s,a) + \delta r_i(s,a), \quad 
    \tilde{\bar r}_i(s) = \bar r_i(s) + \delta \bar r_i(s).
    \]
    Replacing $r_i$ with $\tilde r_i$ for all agents in Eq.~\eqref{eq:logpi_clean} and adjusting the baseline to $\tilde V^*_i(s)$ gives
    \[
    \tilde r_i(s,a) + \frac{\lambda_i}{n-1}\sum_{k\in\group\setminus\{i\}} \tilde{\bar r}_k(s) + \Delta \tilde V^*_i(s,a) 
    = \frac{\log \pi^*_i(s,a)}{\beta} + \delta r_i(s,a) + \frac{\lambda_i}{n-1} \sum_{k\in\group\setminus\{i\}} \delta \bar r_k(s) + \Delta \tilde V^*_i(s,a) - \Delta V^*_i(s,a).
    \]
    For the policies to remain unchanged, we require
    \[
    \delta r_i(s,a) + \frac{\lambda_i}{n-1}\sum_{k\in\group\setminus\{i\}} \delta \bar r_k(s) = \Delta V^*_i(s,a) - \Delta \tilde V^*_i(s,a).
    \]
    Choosing potential-based shaping functions \(\phi_i: \states \to \mathbb{R}\) and setting for all $i$
    \[
    \delta r_i(s,a) = \gamma \sum_{s'} \sum_{\ba_{-i}} \bpi_{-i}(\ba_{-i}\mid s) \transfunc(s'\mid s,\ba) \phi_i(s') - \phi_i(s),
    \]
    and adjusting the baseline as
    \[
    \tilde V^*_i(s) = V^*_i(s) - \Big(\phi_i(s) + \frac{\lambda_i}{n-1}\sum_{k\in\group\setminus\{i\}} \phi_k(s)\Big),
    \]
    ensures that \(\bpi^*\) is invariant.  
    
    \noindent Hence, intrinsic rewards are identifiable only up to potential shaping transformations:
    \[
    \delta r_i(s,a) = \gamma \sum_{s'} \sum_{\ba_{-i}} \bpi_{-i}(\ba_{-i}\mid s) \transfunc(s'\mid s,\ba) \phi_i(s') - \phi_i(s),
    \]
    which proves the proposition.
\end{proof}

\begin{proof}[\textbf{Proof of Corollary}~\ref{th:reduced_ambiguity}]
From Eq.~\eqref{eq:logpi_clean}, the observed QRE policies of agent $i$ in groups $\group$ and $\group'$ satisfy
\begin{align}
    \label{eq:group}
\frac{\log \pi^*_{\group,i}(s,a)}{\beta} &= r_i(s,a) + \frac{\lambda_i}{n-1} \sum_{k \in \group \setminus \{i\}} \bar r_{\group,k}(s) + \Delta V^*_{\group,i}(s,a),\\
\label{eq:group2}
\frac{\log \pi^*_{\group',i}(s,a)}{\beta} &= r_i(s,a) + \frac{\lambda_i}{n-1} \sum_{k \in \group' \setminus \{i\}} \bar r_{\group',k}(s) + \Delta V^*_{\group',i}(s,a).
\end{align}
Any action-dependent shaping $\delta r_i(s,a)$ must be absorbed in both $\Delta V^*_{\group,i}$ and $\Delta V^*_{\group',i}$:
\[
\delta r_i(s,a) = -\delta \Delta V^*_{\group,i}(s,a) = - \delta \Delta V^*_{\group',i}(s,a), \quad \forall a.
\]
Define
\[
\transfunc^{\bpi_{-i}}_a(s) = \mathbb{E}_{\ba_{-i} \sim \bpi_{-i}}[\transfunc(\cdot \mid s, a, \ba_{-i})]
\]
as the transition probabilities induced by the other agents. Then the allowed joint changes in the value functions satisfy
\[
\underbrace{
\begin{pmatrix}
I - \gamma \transfunc^{\bpi_{\group-i}}_{a^1} & I - \gamma \transfunc^{\bpi_{\group'-i}}_{a^1} \\
\vdots & \vdots \\
I - \gamma \transfunc^{\bpi_{\group-i}}_{a^{|\actions|}} & I - \gamma \transfunc^{\bpi_{\group'-i}}_{a^{|\actions|}}
\end{pmatrix}}_{=A}
\begin{pmatrix} V^*_{\group,i} \\ V^*_{\group',i} \end{pmatrix} = 0.
\]
By the rank condition $\operatorname{rank}(A) = 2|\states| - 1$ imposed by our assumptions, the only admissible changes in $V^*$ are constants, so \emph{no action-dependent ambiguity remains} (c.f. \citep{rolland2022identifiability}, proof of Theorem~3).

\noindent For state-only shifts $\delta r_i(s)$, these could in principle be absorbed by other agents' rewards in each group.  
\begin{enumerate}
    \item If $\lambda_i = 0$, changes in other agents' rewards do not affect the right-hand side in Equations~\ref{eq:group} and~\ref{eq:group2}, so $r_i$ is identifiable \emph{up to a constant}.  
    \item If $\lambda_i \neq 0$, state-dependent shifts must be absorbed by the intrinsic rewards of other agents in both groups simultaneously. Denote two other agents $j \in \group$ and $k \in \group'$. The admissible shifts then satisfy
\[
\delta r_i(s) = - \delta \Delta V^*_{\group,j}(s,a) = - \delta \Delta V^*_{\group',k}(s,a), \quad \forall a.
\]
\end{enumerate}
Since we have no guarantee on the rank of the stacked transition matrices for these agents, non-constant shifts in $\delta r_i(s)$ may be \emph{feasible} along directions allowed by the nullspace of the system. Hence, intrinsic rewards can be recovered only up to \emph{non-trivial, state-dependent shifts} when $\lambda_i \neq 0$.  
\end{proof}

\begin{proof}[\textbf{Proof of Corollary}~\ref{th:reward_recovery}]
From Corollary~\ref{th:reduced_ambiguity}, observing each agent $i$ in two groups with the rank condition implies that $r_i$ is recoverable up to state-dependent shifts.

\noindent Now, consider any agent $i$. If $\lambda_i = 0$, state-dependent shifts cannot be absorbed by other agents, so $r_i$ is identifiable \emph{up to a constant}.  

\noindent If $\lambda_i \neq 0$, any state-dependent shift in $r_i$ must be absorbed by other agents' rewards. But each of those agents is also observed in two groups under the rank condition, which similarly restricts their rewards to shifts \emph{only up to a constant}. Consequently, $r_i$ itself can only vary up to a constant.

\noindent Applying this argument to all agents $i \in \{1, \dots, n+1\}$, we conclude that \emph{all intrinsic rewards are identifiable up to a constant}.
\end{proof}

\begin{proof}[\textbf{Proof of Theorem~\ref{th:full_recovery}}]
We want to show that the altruism levels $\lambda_i$ are uniquely identifiable and that the intrinsic rewards $r_i$ can be recovered up to a constant.

\noindent Consider an arbitrary agent $i$, observed in two groups $\group$ and $\group'$ satisfying the rank condition from Corollary~\ref{th:reduced_ambiguity}. Suppose we perturb the altruism of agent $i$ by $\delta \lambda_i$, i.e., $\tilde \lambda_i = \lambda_i + \delta \lambda_i$, such that it does not change the observed QRE policies. We have, for instance in group $\group$:

\begin{equation}
    \frac{\log \pi^*_{\group, i}(s,a)}{\beta} = r_i(s,a) + \frac{\tilde \lambda_i}{n-1}\sum_{k\in\group\setminus\{i\}} \bar r_k(s) + \Delta V^*_i(s,a),
\end{equation}
To maintain the same observed QRE policies, one could attempt to compensate for this perturbation by:
\begin{enumerate}
    \item Adjusting the value function of agent $i$ in each group,  
    \item Adjusting agent $i$'s intrinsic reward $r_i(s,a)$, or  
    \item Adjusting the intrinsic rewards of other agents $r_k(s,a), k \neq i$.
\end{enumerate}

\medskip

\noindent \textbf{Option (1): Adjusting $V^*_i$}  

\noindent From Corollary~\ref{th:reward_recovery}, $V^*_{\group,i}$ can only change by a constant without affecting the policy. But the altruism perturbation introduces a state-action dependent term:
\[
\frac{\delta \lambda_i}{n-1} \sum_{k \in \group \setminus \{i\}} {\bar r}_{\group,k}(s),
\]
which \emph{cannot} be constant, unless $\delta \lambda_i = 0$. Therefore, option (1) is discarded.

\medskip

\noindent \textbf{Option (2): Adjusting $r_i(s,a)$}  

\noindent To offset $\delta \lambda_i$, we would need to perturb the intrinsic rewards with
\[
\tilde r_i(s,a) = r_i(s,a) - \frac{\delta \lambda_i}{n-1} \sum_{k \in \group \setminus \{i\}} {\bar r}_{\group,k}(s), \quad \forall \group \;\text{containing}\; i,
\]  
for all groups in which $i$ appears. This requires the quantity $\sum_{k \in \group \setminus \{i\}} {\bar r}_{\group,k}(s)$ to be constant across groups containing $i$.
Recalling that $\bar r_{\group, k} = \sum_{a_k} \bpi^*_{\group, k}(a_k\mid s) r_k(s, a)$, this must imply that rewards and policies are identical across agents and groups.
However, because we know groups $\group$ and $\group'$ verify the rank condition on their induced transition functions, the other policies cannot be  identical $\bpi_{\group-i} \neq \bpi_{\group'-i}$.
Hence, option (2) fails.

\medskip

\noindent \textbf{Option (3): Adjusting $r_k(s,a), k \in \group \cup \group' \setminus \{i\}$}  

\noindent If $\lambda_i = 0$, other agents' rewards do not influence agent $i$'s policy, so no compensation is possible.  

\noindent If $\lambda_i \neq 0$, compensating requires rescaling $\bar r_{\group,k}(s)$ across groups. 
Starting from the equilibrium condition, we must have:
\begin{align*}
    \lambda_i \tilde\bar r_{\group,k}(s) &= \lambda_i \bar r_{\group,k}(s) - \delta \lambda_i \tilde \bar r_{\group,k}(s)\\
    \tilde \lambda_i \tilde\bar r_{\group,k}(s) &= \lambda_i \bar r_{\group,k}(s)\\
    \tilde\bar r_{\group,k}(s) &= \frac{\lambda_i}{\tilde \lambda_i} \bar r_{\group,k}(s),
\end{align*}
where the last equality assumes $\tilde \lambda_i \neq 0$. Thus, any compensation for a change in altruism for agent $i$ necessarily rescales the expected rewards of the other agents by the factor $\frac{\lambda_i}{\tilde \lambda_i}$.
Since each expected reward satisfies
\[
    \bar r_{\group, k}(s, a) = \sum_a \pi^*_{\group, k}(a \mid s) r_k(s,a),
\] 
and the observed policy $\pi^*_{\group, k}$ is fixed, the only way to realise this rescaling is to multiply every intrinsic reward $r_k$ by the same factor $\frac{\lambda_i}{\tilde \lambda_i}$.

\noindent Now consider some specific agent $k \neq i$, whose intrinsic reward has been rescaled by this factor. In any observed $\group''$ that includes agent $k$, we have
    \[
        \frac{\log \pi^*_{\group'', k}(s, a)}{\beta} = \frac{\lambda_i}{\tilde \lambda_i} r_k(s, a) + \frac{\lambda_k}{n-1}\sum_{\substack{j\in\group'' \\ j \neq k}} \bar r_{\group'', j}(s) + \Delta V^*_{\group'', k}(s, a),
    \]
This rescaling introduces a multiplicative perturbation term $\frac{\delta \lambda_i}{\lambda_i + \delta \lambda_i}r_k(s, a)$, which both depends on actions and states. To maintain identical QRE policies, this term would need to be absorbed by transforming its value differences $\Delta V^*_{\group'', k}$. However, because there exists at least one pair of observed groups including $k$ such that verify the rank condition from Corollary~\ref{th:reduced_ambiguity}, any valid compensation could only correspond to a constant shift. Yet, the perturbation $\frac{\delta \lambda_i}{\lambda_i + \delta \lambda_i}r_k(s, a)$ is not constant across states or actions. Therefore, such compensation is impossible, and option (3) also fails.  

\medskip

\noindent \textbf{Conclusion on $\lambda_i$}  

\noindent No combination of adjustments can absorb a nonzero $\delta \lambda_i$. Therefore, each altruism parameter $\lambda_i$ is uniquely determined by the observed policies. That is, $\lambda_1, \dots, \lambda_{n+1}$ are \emph{exactly identifiable}, independently of the intrinsic rewards.

\medskip

\noindent \textbf{Recovery of intrinsic rewards}  

\noindent Once altruism levels are disentangled, Corollary~\ref{th:reward_recovery} implies that the intrinsic rewards $r_i$ can be recovered \emph{up to constants}, completing the proof.
\end{proof}

\newpage

\section{Algorithms}

\label{appendix:algorithms}

\begin{algorithm}[h]
\caption{DRP-SGLD (Direct Reward Posterior Sampling)}\label{algo:drp}
\begin{algorithmic}[1]
\STATE \textbf{Input:} Priors $\bel(\Rset)$, $\bel(\beta)$; demonstrations $\demosall$; number of iterations $T$; warm-up $w<T$; step size $\epsilon_t$.
\STATE \textbf{Initialise:} Reward parameters $\psi_0 \sim \bel(\psi)$.
\FOR{$t = 0, \dots, T-1$}
    \STATE Initialize gradient accumulator $g_t \leftarrow 0$
    \FOR{each group $\group$ in $\demosall$}
        \STATE Sample $\beta_t \sim \bel(\beta)$
        \STATE Compute equilibrium policy $\bpi^*_{\bR_{\psi_{t}, \group}}(\cdot; \beta_t)$
        \STATE Accumulate gradient: 
        \[
            g_t \mathrel{+}= 
            \nabla_\psi \sum_{\tau \in \demos_\group} 
            \sum_{(s, \ba) \in \tau} 
            \log \bpi^*_{\bR_{\psi_{t}, \group}}(\ba \mid s; \beta_t)
        \]
    \ENDFOR
    \STATE Sample noise $\eta_t \sim \mathcal{N}(0, \epsilon_t)$
    \STATE Update parameters:
    \[
        \psi_{t+1} \leftarrow \psi_{t} + 
        \frac{\epsilon_t}{2} 
        \left( 
            \nabla_\psi \log \bel(\Rset) + g_t 
        \right) 
        + \eta_t
    \]
\ENDFOR
\STATE \textbf{Return:} Posterior samples $\left(\Rset_{\psi_t}\right)_{t=w}^T$
\end{algorithmic}
\end{algorithm}

\begin{algorithm}[h]
\caption{PORP-SGLD (Policy Oriented Reward Posterior Sampling)}\label{algo:porp}
\begin{algorithmic}[1]
\STATE \textbf{Input:} Priors $\bel(\Rset)$, $\bel(\beta)$, $\bel(\pi)$; gap function $\Delta$; optimality concentration $c$; demonstrations $\demosall$; 
number of iterations $(T^\pi, T^R)$; warm-ups $(w^\pi<T^\pi, w^\pi<T^R)$; step sizes $(\epsilon^\pi_t; \epsilon^R_t)$.

\STATE \textbf{1. Policy sampling step}
\FOR{each group $\group$ in $\demosall$}
    \STATE \textbf{Initialise:} Policy parameters for group $\group$: $\theta_0 \sim \bel(\theta)$.
    \FOR{$t = 0, \dots, T^\pi-1$}
        \STATE Compute gradient:
        \[
            g_t = \nabla_\theta \sum_{\tau \in \demos_\group}\sum_{(s, \ba) \in \tau} \log \bpi_{\theta_{t}}(\ba \mid s)
        \]
    \STATE Sample noise $\eta_t \sim \mathcal{N}(0, \epsilon^\pi_t)$
    \STATE Update parameters:
    \[
        \theta_{t+1} \leftarrow \theta_{t} + 
        \frac{\epsilon^\pi_t}{2} 
        \left( 
            \nabla_\theta \log \bel(\bpi_{\theta_{t}}) + g_t 
        \right) 
        + \eta_t
    \]
    \ENDFOR
    \STATE Store policy samples for group $\group$: $\mathbf{P}_\group \leftarrow \left(\bpi_{\theta_{t}}\right)_{t=w^\pi}^{T^\pi}$
\ENDFOR

\STATE \textbf{2. Reward sampling step}
\STATE \textbf{Initialise:} Reward parameters $\psi_0 \sim \bel(\psi)$.

\FOR{$t = 0, \dots, T^R-1$}
    \STATE Initialise gradient accumulator $g_t \leftarrow 0$
    \FOR{each group $\group$ in $\demosall$}
        \STATE Sample $\bpi \sim \mathcal{U}(\mathbf{P}_\group)$
        \STATE Sample $\beta_t \sim \bel(\beta)$
        \STATE Accumulate gradient: 
        \[
            g_t \mathrel{-}= 
            c \nabla_\psi  \Delta_{\bR_{\psi_t, \group},\beta_t} \left(\bpi \right)
        \]
    \ENDFOR
    \STATE Sample noise $\eta_t \sim \mathcal{N}(0, \epsilon^R_t)$
    \STATE Update parameters:
    \[
        \psi_{t+1} \leftarrow \psi_{t} + 
        \frac{\epsilon_t}{2} 
        \left( 
            \nabla_\psi \log \bel(\Rset) + g_t 
        \right) 
        + \eta_t
    \]
\ENDFOR
\STATE \textbf{Return:} Reward posterior samples $\left(\Rset_{\psi_t}\right)_{t=w^R}^{T^R}$
\end{algorithmic}
\end{algorithm}

\subsection{DRP and PORP implementation}

We implement sampling from both the DRP and PORP using Stochastic Gradient Langevin Dynamics (SGLD), as shown in Algorithms~\ref{algo:drp} and~\ref{algo:porp}. Reward parameters are shared across groups that include the same agents, we give more details on parametrisation in Appendix~\ref{sec:parametrisation}. All priors are flexible; the specific choices used in our experiments are listed in Table~\ref{tab:priors}.

In standard SGLD, the gradient of the log-posterior is scaled by the minibatch fraction $\frac{b}{B}$, with $b$ the minibatch size and $B$ the dataset size. Here, we treat $\beta$ as a resampled variable, effectively emulating an infinite dataset. We observed that omitting this scaling and tempering the step size yields consistent results.

We adopt an RMSProp-based variant of SGLD~\citep{li2016preconditioned} with momentum $0.99$ and $\epsilon = 10^{-8}$. Step sizes decay according to $\epsilon_t = \epsilon_0 / (1+t)^\alpha$, with the hyperparameters summarised in Table~\ref{table:sgld_hypers}.
\subsection{SGLD for Entropy Regularised IRL}

We employ SGLD for its efficiency, scalability, and rapid convergence in high-dimensional settings. Crucially, all of our proposed posteriors, over policies and rewards for both DRP and PORP, are log-differentiable, enabling efficient gradient computation of the log-posterior and straightforward application of SGLD updates. Moreover, the ability to perform minibatched inference further enhances scalability, particularly when datasets are large or evaluating even a single sample is memory-intensive.

SGLD-based inference is particularly well-suited to entropy-regularised reinforcement learning, where the dependence of policies on rewards is smooth: small changes in rewards induce smooth changes in policies. As a result, log-posteriors over rewards and policies inherits this smoothness, creating a favorable landscape for gradient-based sampling.

While we do not provide formal convergence proofs for SGLD in our setting, we note that with a Gaussian prior over rewards, the DRP and PORP reward posteriors (both PSG and QIG variants) become strongly concave, ensuring rapid convergence of SGLD~\citep{raginsky2017non, zou2021faster}. Our additional results confirm this observation, where we observe that it is hard to infer anything on weak Gaussian priors~\ref{fig:reward_prior}. For PORP in particular, one can also interpret Eq.~\ref{eq:porp} as a mixture of log-concave distributions, suggesting that the overall posterior may satisfy log-Sobolev properties, which are known to support fast mixing and convergence in high dimensions~\citep{vempala2019rapid}.

Other standard MCMC methods, such as Gibbs sampling, Metropolis-Hastings, or policy-walk approaches, are prohibitively slow and inefficient in our setting due to the high dimensionality of joint policies and rewards. SGLD leverages gradient information to achieve many orders-of-magnitude faster mixing.

\section{Additional experimental results}
\label{apx:results}

\subsection{Constant nature of $Z_\bpi$}

The normalising constant $Z_\bpi$ plays a crucial role in PORP sampling (Eq.~\eqref{eq:porp}). If $Z_\bpi$ varies significantly across policies, sampling becomes inefficient. To address this, we justify approximating $Z_\bpi \approx Z$ through both theoretical and empirical analyses.

$Z_\bpi$ arises from the posterior over joint rewards given a joint policy and entropy parameter (Eq.~\eqref{eq:gap_posterior}), and depends on the chosen gap function. Here, we analyse two gap functions: QIG and PSG. Throughout, we fix an arbitrary group of agents and a value of $\beta$, dropping corresponding subscripts for clarity.

\noindent \textbf{Theoretical analysis}

\paragraph{\textbf{QIG}}
We approximate $Z_\bpi$ using the Laplace approximation. In the entropy-regularised setting, there is a unique optimal policy for a given reward function, so there exists a unique reward $\bR^*$ such that $\bpi$ is optimal: $\bpi = \bpi^*_{\bR^*}$. By definition of QIG, this implies $\gapv_{\bR^*}(\bpi) = 0$.

Let $H_\bpi$ denote the negative Hessian of the log posterior at $\bR^*$:
\[
H_\bpi = \left. \nabla^2_{\bR} c \, \gapv_{\bR}(\bpi) \right|_{\bR = \bR^*}.
\]
If $H_\bpi$ is positive semidefinite, the Laplace approximation yields:
\[
Z_\bpi \approx \exp\left(\gapv_{\bR^*}(\bpi)\right) \sqrt{\frac{(2\pi_{\text{geom}})^{|\states||\actions|}}{|H_\bpi|}}
= \sqrt{\frac{(2\bpi_{\text{geom}})^{|\states||\actions|}}{|H_\bpi|}},
\]
where $\pi_{\text{geom}}$ is the geometric constant.

We now argue that $H_\bpi$ is both positive semidefinite and independent of $\bpi$. Dropping the concentration factor $c$ and summing over states, the first derivative w.r.t.\ rewards for the player $i$ maximizing the gap at $\bR^*$ is:
\[
\nabla_\bR V_i^* - V_i^\bpi = \nabla_\bR V_i^* - \left(\mathbf{I} - \gamma \mathbf{\transfunc}^{\bpi}\right)^{-1}.
\]
Since $V_i^\bpi$ is linear in $\bR$ and $V_i^*$ is the pointwise maximum over policies, $V_i^*$ is convex in $\bR$. Consequently, the Hessian $H^\bpi$ is independent of $\bpi$ and positive semidefinite.

Thus, for QIG, we conclude
\[
Z_\pi \approx Z, \;\text{with}\; Z \approx \sqrt{\frac{(2\pi_{\text{geom}})^{|\states||\actions|}}{|H|}}.
\]

\paragraph{\textbf{PSG}}
Consider a set of sampled policies $(\bpi_k)_k$ from the first step of PORP. By construction, each $\bpi_k$ lies in an $\epsilon$-ball around the true policy $\bpi$:
\[
\bpi_k \in \left\{ \dkl(\bpi(\cdot \mid s) | \bpi) < \epsilon, \forall s \right\}.
\]
As more demonstrations are gathered, $\epsilon$ shrinks, and sampled policies become closer to $\bpi$. We want to bound
\[
\frac{Z_{\bpi_j}}{Z_{\bpi_k}} = \frac{\int e^{-c \gapkl_\bR(\bpi_j)} , d\bR}{\int e^{-c \gapkl_\bR(\bpi_k)} , d\bR}.
\]

Denote $\Delta_\text{KL}(\bR) := \gapkl_\bR(\bpi_j) - \gapkl_\bR(\bpi_k)$. Then
\[
Z_{\bpi_j} = \int e^{-c \gapkl_\bR(\bpi_k)} e^{-c \Delta_\text{KL}(\bR)} , d\bR,
\]
so that
\[
\frac{Z_{\bpi_j}}{Z_{\bpi_k}} = \mathbb{E}_{\bR \sim \bel(\cdot \mid \bpi_k)} \big[e^{-c \Delta_\text{KL}(\bR)}\big].
\]

Assume $|\Delta_\text{KL}(\bR)| \le \delta$ for all $\bR$. Applying Hoeffding's lemma:
\[
\log \mathbb{E}[e^{-c\Delta_\text{KL}(\bR)}] \in \Big[-c\,\mathbb{E}[\Delta_\text{KL}(\bR)] - \frac{c^2\delta^2}{2}\, \; -c\,\mathbb{E}[\Delta_\text{KL}(\bR)] + \frac{c^2\delta^2}{2}\Big].
\]
Exponentiating gives
\[
e^{-c^2 \delta^2/2 - \kappa_{\bpi_k}} \le \frac{Z_{\bpi_j}}{Z_{\bpi_k}} \le e^{c^2 \delta^2/2 - \kappa_{\bpi_k}}, \quad
\kappa_{\bpi_k} = c\, \mathbb{E}_{\bR \sim \bel(\cdot \mid \bpi_k)}[\Delta_\text{KL}(\bR)].
\]

Next, we relate $\delta$ to $\epsilon$ using the fact that sampled policies are $\epsilon$-close to $\bpi$. Bounding the entropy difference using the Fannes-Audenaert inequality and the softmax difference using Lipschitz properties of $\log$-softmax and $\bar Q$, we obtain
\[
|\Delta_\text{KL}(\bR)| \le |\states| \Big( 2\sqrt{2\epsilon} C + \ent_2(2\sqrt{2\epsilon}) \Big),
\]
with
\[
C := \log(|\actions|-1) + \frac{2||R||_\infty + \frac{1}{\beta} \log |\actions|}{1 - \gamma} + \log |\actions| + 2\beta K.
\]

Similarly,
\[
\kappa_{\bpi_k} \le c |\states| \Big( 2\sqrt{2\epsilon} C + \ent_2(2\sqrt{2\epsilon}) \Big).
\]

Combining these bounds gives
\[
\exp\Big(-|\states|(d + \frac{c^2 |\states| d^2}{2})\Big) \le \frac{Z_{\bpi_j}}{Z_{\bpi_k}} \le \exp\Big(|\states|(d + \frac{c^2 |\states| d^2}{2})\Big), \quad
d := 2\sqrt{2\epsilon} C + \ent_2(2\sqrt{2\epsilon}).
\]

As $\epsilon \to 0$, this implies
\[
\frac{Z_{\bpi_j}}{Z_{\bpi_k}} \approx 1 \pm \mathcal{O}(|\states| \sqrt{\epsilon} |\log \epsilon|),
\]
showing that $Z_\bpi \approx Z$ asymptotically. Intuitively, more demonstrations reduce $\epsilon$, leading to nearly constant $Z_\bpi$. However, the convergence rate $\sqrt{\epsilon} |\log \epsilon|$ is relatively slow, and $\epsilon$ may remain non-negligible in practical settings. To verify that $Z_\bpi$ nevertheless remains approximately constant, we next provide an empirical evaluation.

\noindent \textbf{Empirical analysis}

To complement our theoretical results, we conducted an empirical study to investigate the invariance of the normalising factor $Z_\bpi$ across different policies, for both the QIG and PSG gap functions. We considered Markov games (MGs) of varying state and action space sizes and sampled 200 policies from the policy posterior, corresponding to the first step of the PORP procedure. To show that $Z_\bpi \approx Z$ even in low demonstration regimes, we experiment with 3 and 36 trajectories of length 100 to compute the policy posterior. For each sampled policy $\bpi$, we then estimated the corresponding value of $Z_\bpi$.

Because the gap distributions are highly concentrated around the ground truth rewards $\bR = (\bl, \br)$, we generated $N = 2 \times 10^4$ samples of altruism levels and intrinsic rewards from truncated normal distributions with small variance:
\[
\hat \bl_i \sim \mathcal{N}_{[\lambda_\text{min}, \lambda_\text{max}]}(\bl, 0.16), \quad \hat \br_i \sim \mathcal{N}_{[r_\text{min}, r_\text{max}]}(\br, 0.16).
\]

We then estimated $Z_\bpi$ via importance sampling:
\[
\hat{Z}_\bpi = \frac{1}{N} \sum_{i=1}^N e^{\Delta_{(\hat \bl_i, \hat \br_i)}(\bpi)} \cdot (r_\text{max} - r_\text{min}) (\lambda_\text{max} - \lambda_\text{min}) \, q(\hat \bl_i, \hat \br_i),
\]
where $q(\hat \bl_i, \hat \br_i)$ is the density of the joint truncated normal distribution evaluated at $(\hat \bl_i, \hat \br_i)$.

\begin{figure}
    \centering
    \begin{subfigure}[t]{0.49\linewidth}
    \includegraphics[width=\linewidth]{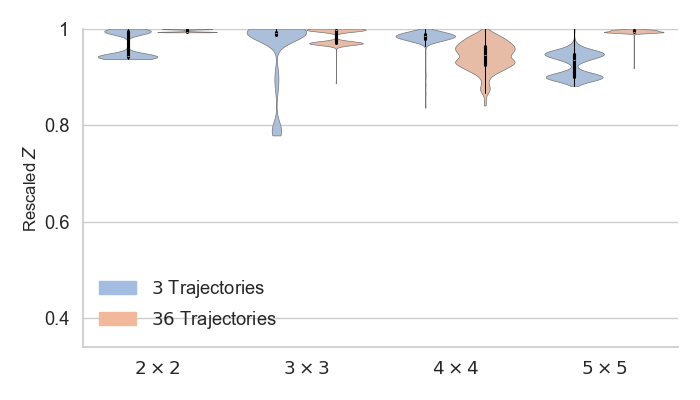}
    \caption{QIG.}
    \end{subfigure}
    \begin{subfigure}[t]{0.49\linewidth}
    \includegraphics[width=\linewidth]{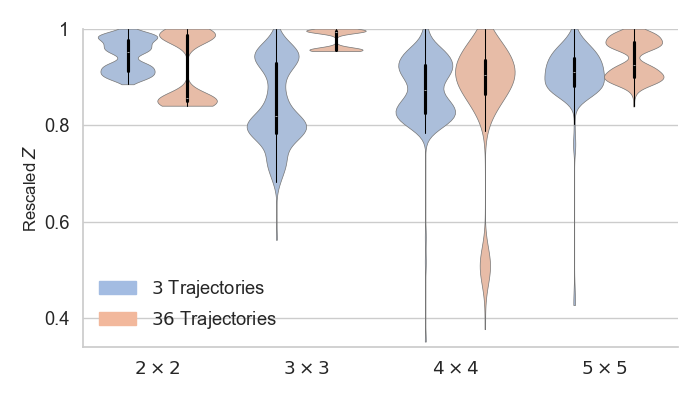}
    \caption{PSG.}
    \end{subfigure}
    \caption{Rescaled normalising constant $Z_\bpi$ estimated across random games of varying sizes ($|\states| \times |\actions|$), gap functions, and 200 policies sampled from the policy posterior step of the PORP. Each value of $Z_\bpi$ is estimated using $2 \times 10^4$ reward samples and rescaled by the maximum value attained across all policies.}
    \label{fig:z_plot}
\end{figure}

We report the results in Figure~\ref{fig:z_plot}, where we set $r_\text{min} = 0, r_\text{max} = 1, \lambda_\text{min} = 0,$ and $\lambda_\text{max} = 1$. Our results show that $Z_\bpi$ remains nearly constant across different policy deployments, with variations typically within $20\%$ for both gap functions. This empirical evidence supports the theoretical conclusion that $Z_\bpi \approx Z$, justifying the use of a constant normalising factor in the PORP procedure.

\subsection{Additional results in Random MGs} 

\begin{figure}[t]
    \centering
    \begin{subfigure}[b]{0.45\linewidth}
        \centering
        \includegraphics[width=\linewidth]{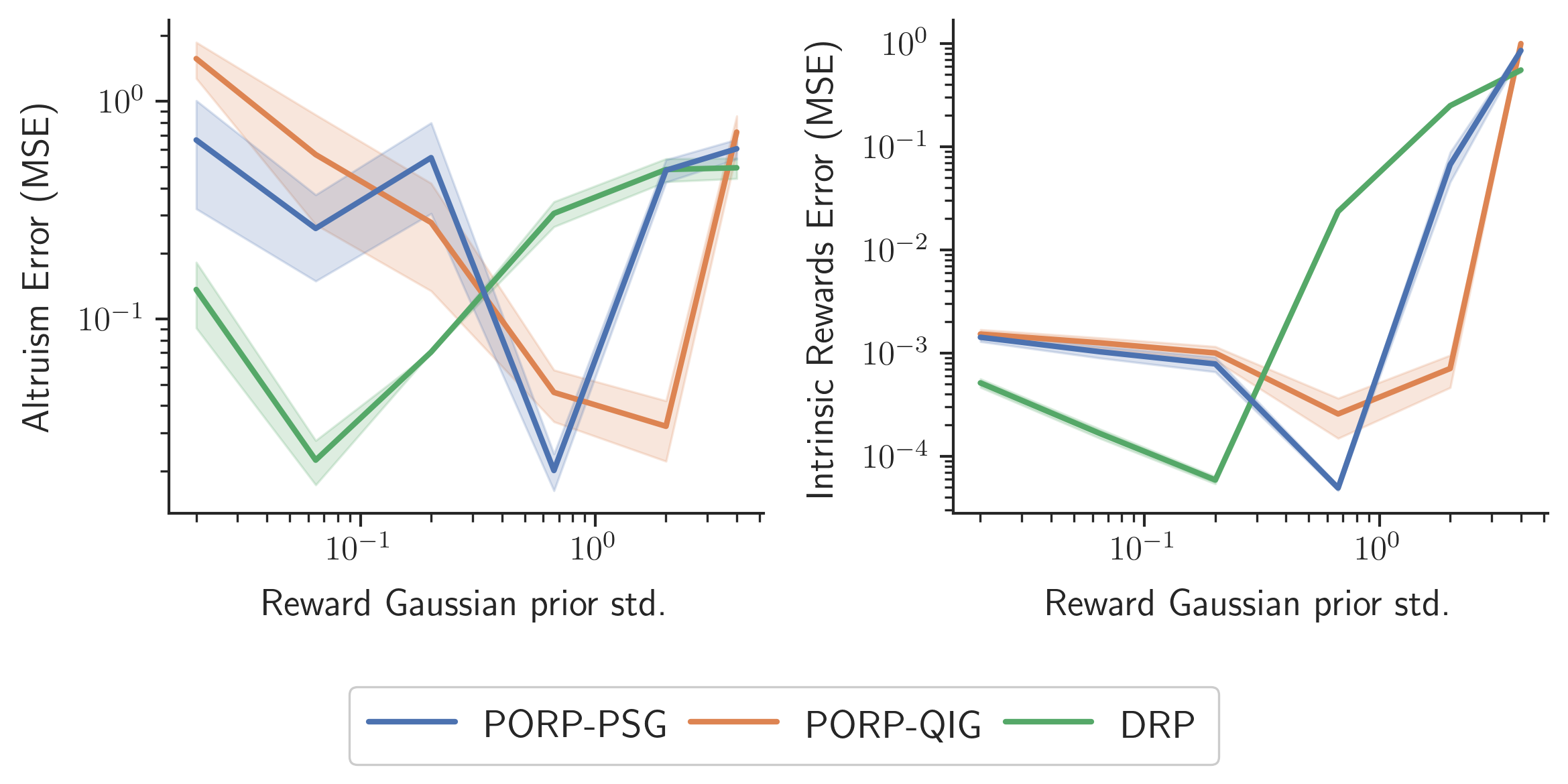}
        \caption{Error over reward prior strength.}
        \label{fig:reward_prior}
    \end{subfigure}
    \begin{subfigure}[b]{0.45\linewidth}
        \centering
        \includegraphics[width=\linewidth]{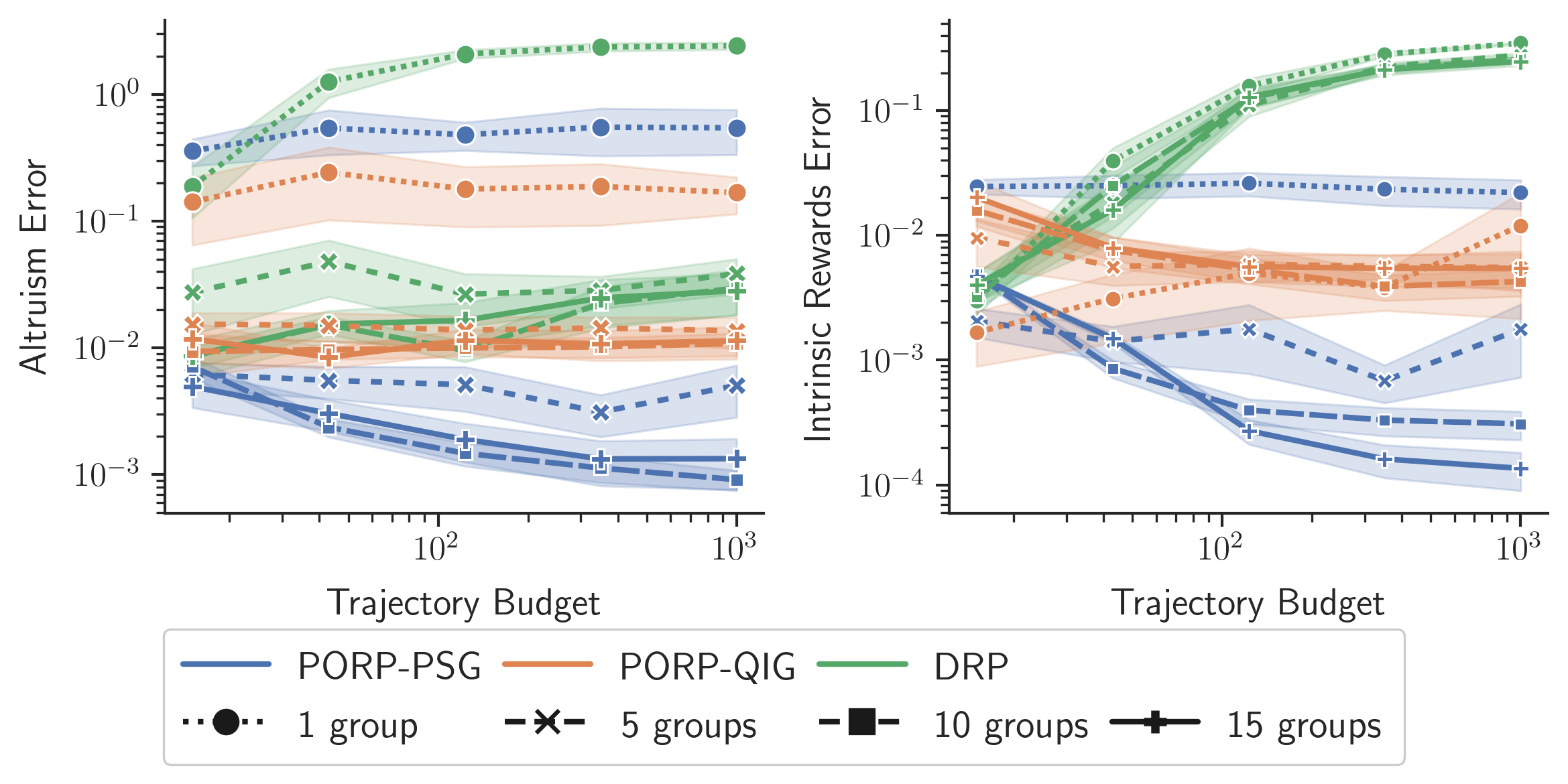}
        \caption{Error using different numbers of groups and demonstrations.}
        \label{fig:se}
    \end{subfigure}
    \caption{Effect of reward Gaussian priors on inference, and sample efficiency of methods with increasing number of demonstrations and groups.}
    \label{fig:priors}
\end{figure}

\paragraph{\textbf{Sample efficiency with groups}}
We include the standard error of our methods under different number of trajectories and group sizes in Figure~\ref{fig:se}.

\paragraph{\textbf{Prior robustness}}

We conducted additional experiments on randomly generated MGs to investigate how different prior assumptions and demonstration optimalities affect inference. All metrics were averaged over 5 random seeds and computed on MGs with 343 states, 5 actions, and 3 players.

\begin{figure}[t]
    \centering
    \begin{subfigure}[b]{0.45\linewidth}
        \centering
        \includegraphics[width=\linewidth]{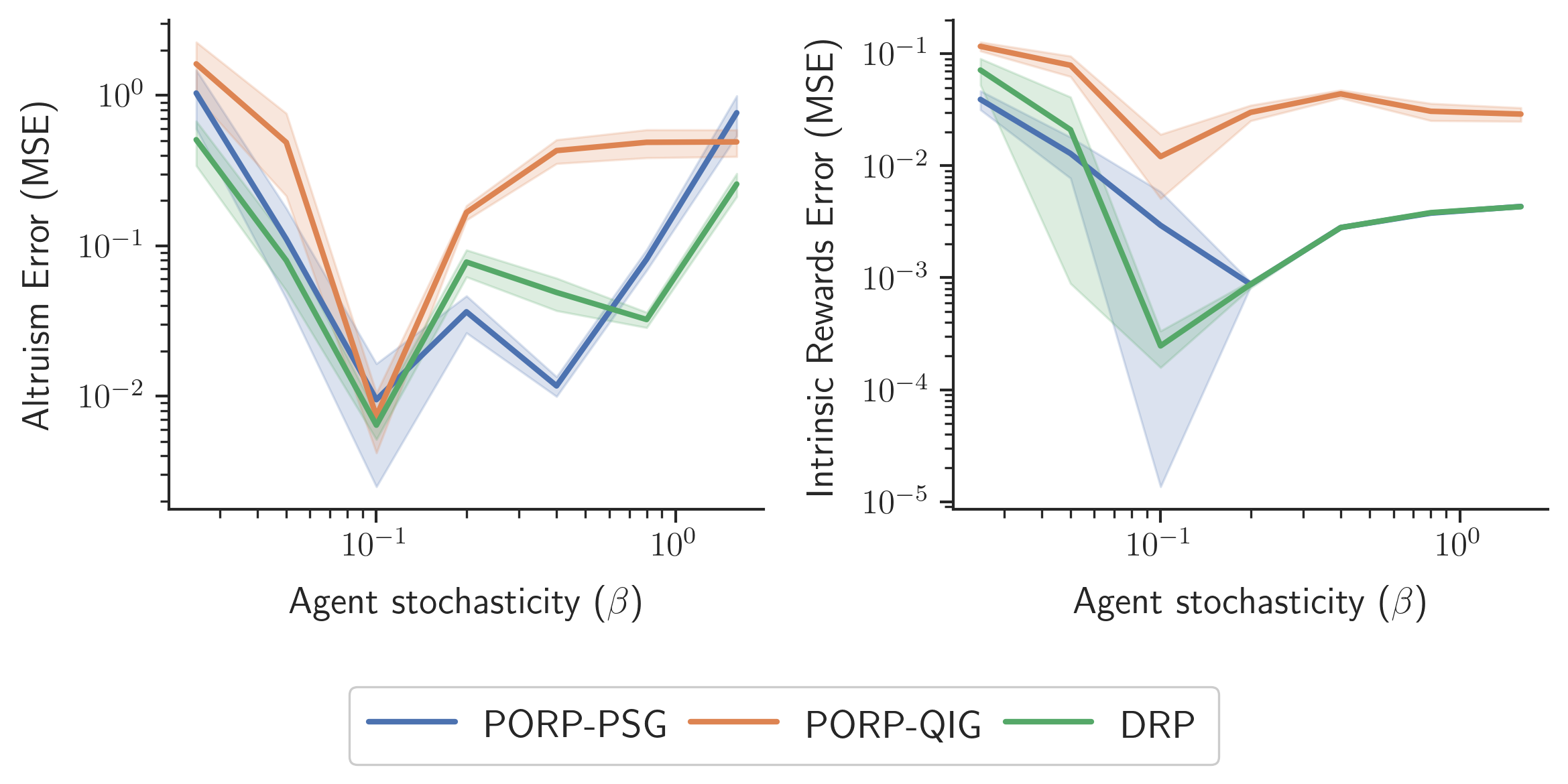}
        \caption{Error over demonstration stochasticity.}
        \label{fig:true_beta}
    \end{subfigure}
    \begin{subfigure}[b]{0.45\linewidth}
        \centering
        \includegraphics[width=\linewidth]{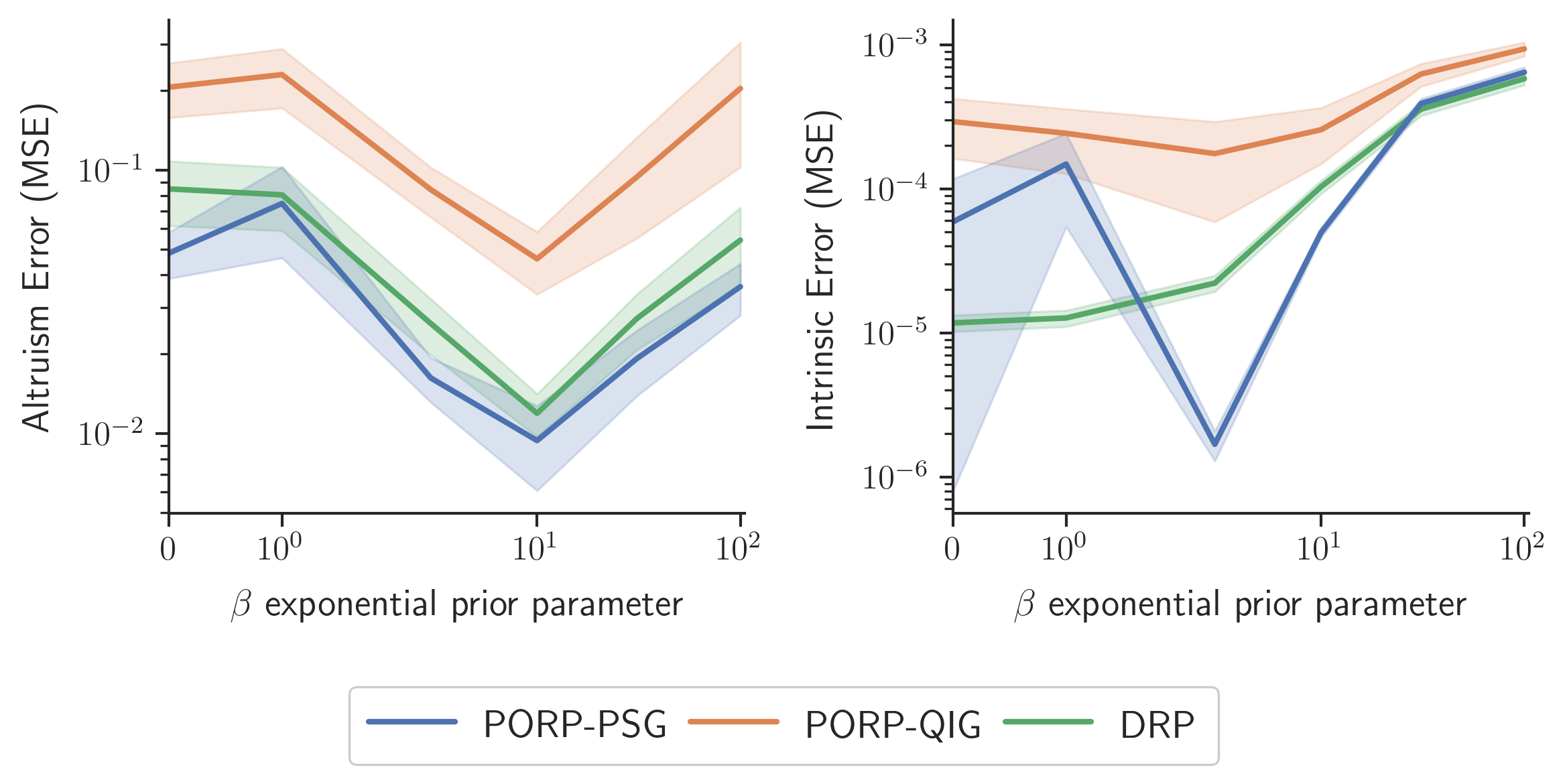}
        \caption{Error over agent stochasticity belief.}
        \label{fig:beta_prior}
    \end{subfigure}
    \caption{Robustness of methods over varying optimality of demonstrations, and stochasticity belief.}
    \label{fig:optimality}
\end{figure}

In Figure \ref{fig:true_beta}, we vary the true latent optimality parameter $\beta$ to study how DRP and PORP perform when demonstrations range from highly stochastic to nearly optimal (in an unregularised sense). The results reveal that inference becomes difficult at both extremes, neither overly random nor fully deterministic behaviour is easy to interpret. Notably, PORP-PSG exhibits stable performance for a broad range of $\beta$ values between 0.05 and 0.5.

Figure \ref{fig:beta_prior} examines robustness with respect to our belief about demonstration optimality, transitioning from a uniform prior to one sharply peaked around $\beta_\text{min}$. Both DRP and PORP-PSG remain largely insensitive to this shift, achieving accurate intrinsic reward estimates even when the prior is completely uniform.

Finally, Figure \ref{fig:reward_prior} explores how inference behaves as the prior over rewards varies in strength. When the Gaussian reward prior is either too tight (small standard deviation) or too loose (nearly uniform), both methods struggle to recover altruism and intrinsic rewards. In the extreme case of a near-uniform prior, intrinsic reward estimation collapses to random guessing.

Overall, these results suggest that while both DRP and PORP-PSG exhibit strong resilience to prior misspecification, inferring altruism remains more sensitive to hyperparameter choices than estimating intrinsic rewards.

\section{Additional experimental details}
\label{apx:details}

\subsection{Policy and reward parametrisation} 
\label{sec:parametrisation}

We parameterise policies using a softmax over logits. For state $s$ and action $a$, the policy is
\[
\pi_{\boldsymbol{\theta}}(s, a) = \frac{\exp(\theta_{s, a})}{\sum_{a'} \exp(\theta_{s, a'})},
\]
with $\boldsymbol{\theta} \in \mathbb{R}^{|\mathcal{S}| \times |\mathcal{A}|}$ representing the policy parameters. SGLD directly operates on this parameter space.

For rewards, each agent $i$ has its own intrinsic reward parameters $\psi^r_i$ and altruism parameter $\psi^\lambda_i$. Intrinsic rewards are scaled via an offset sigmoid:
\[
r_i(s, a) = \text{sigmoid}(\psi^r_{i,s,a}) \cdot (r_\text{max} - r_\text{min}) + r_\text{min},
\]
and altruism is similarly scaled:
\[
\lambda_i = \text{sigmoid}(\psi^\lambda_i) \cdot (\lambda_\text{max} - \lambda_\text{min}) + \lambda_\text{min}.
\]
By mapping reward and altruism parameters through a sigmoid, we ensure bounded outputs and gradients, improving stability during training.
In our experiments, all reward parameters $\psi^r_i$ and $\psi^\lambda_i$ are clipped to $[-9, 9]$ to avoid saturation of the sigmoid.

\subsection{Random MGs parameters} 

For all experiments, we used a discount factor of $\gamma=0.9$, and an entropy parameter of $\beta=0.1$. To generate transition probabilities in randomised Markov games, we used a Dirichlet distribution with parameter $0.3$.

\subsection{Overcooked} 

\begin{figure}[t]
    \centering
    \adjustbox{max width=\textwidth}{
    \begin{tabular}{cccccc}
        \subcaptionbox{Step 0}{\includegraphics[trim={0 0 0 2.4cm},clip,width=5cm]{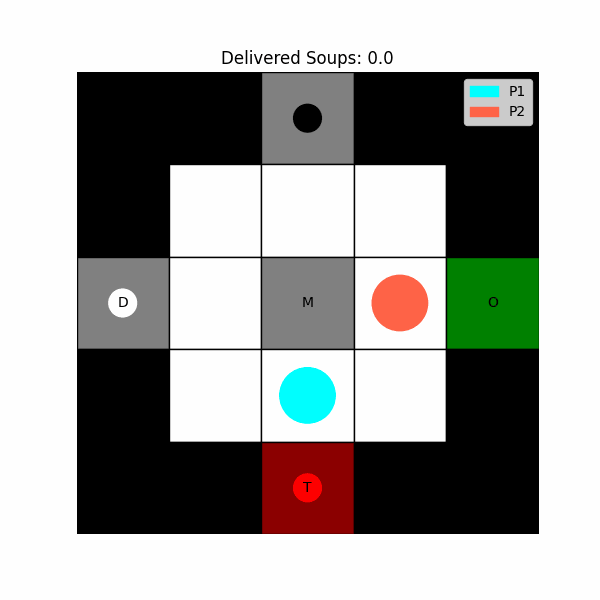}} &
        \subcaptionbox{Step 1}{\includegraphics[trim={0 0 0 2.4cm},clip,width=5cm]{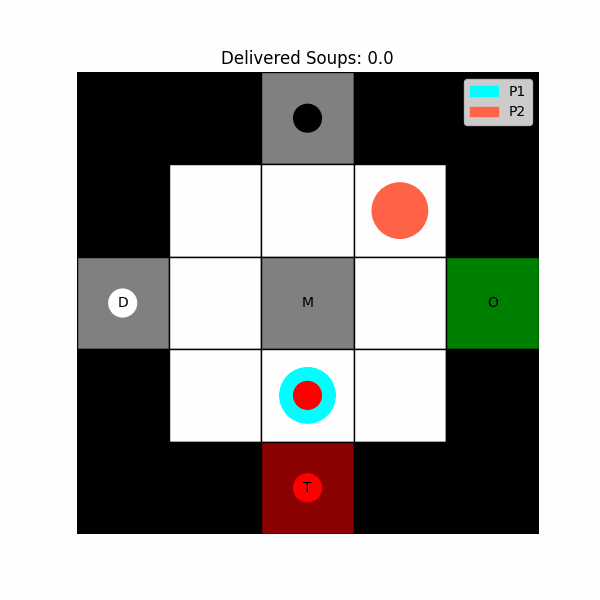}} &
        \subcaptionbox{Step 2}{\includegraphics[trim={0 0 0 2.4cm},clip,width=5cm]{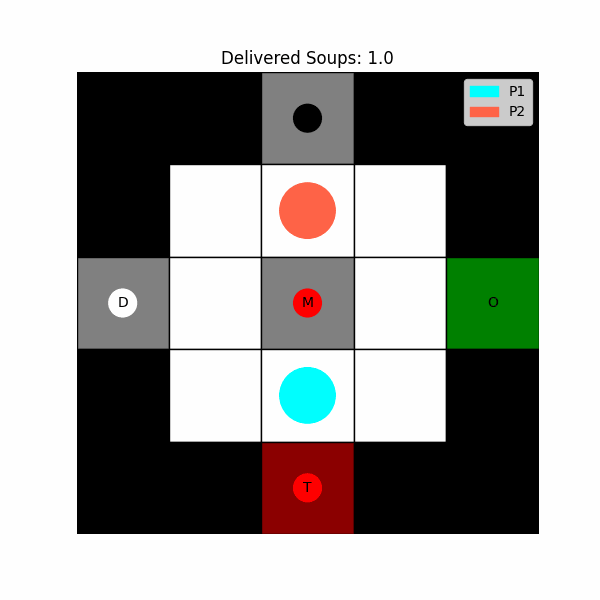}} &
        \subcaptionbox{Step 3}{\includegraphics[trim={0 0 0 2.4cm},clip,width=5cm]{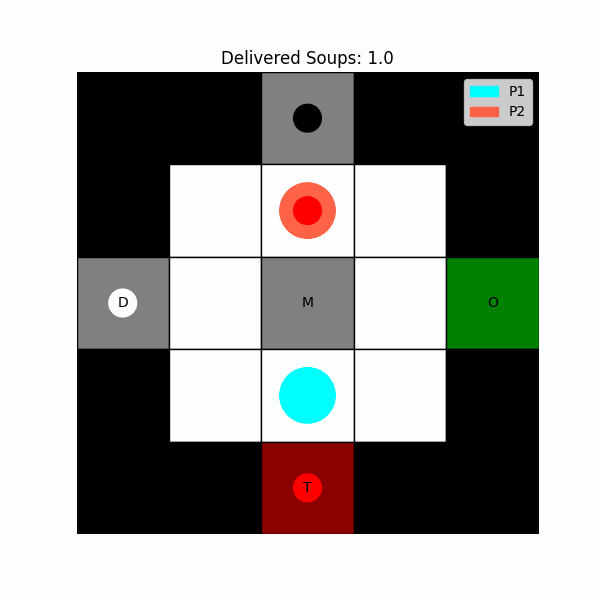}} &
        \subcaptionbox{Step 4}{\includegraphics[trim={0 0 0 2.4cm},clip,width=5cm]{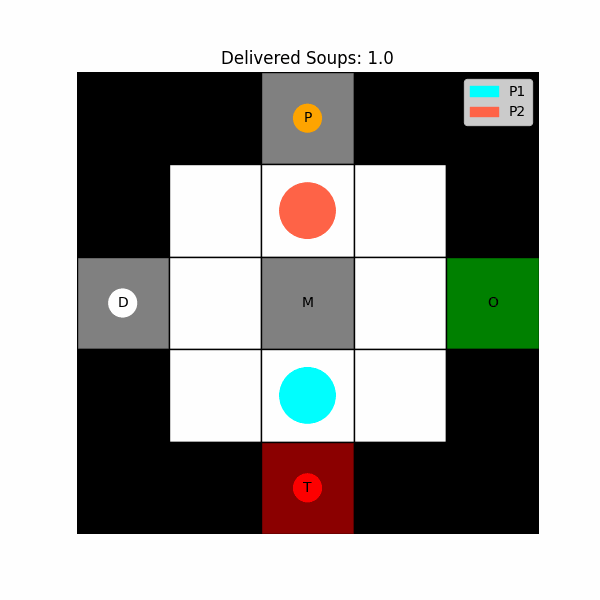}} \\
        \subcaptionbox{Step 5}{\includegraphics[trim={0 0 0 2.4cm},clip,width=5cm]{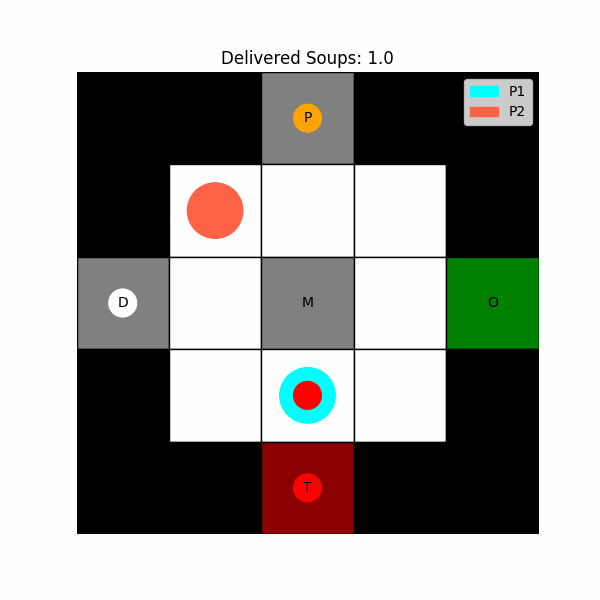}} &
        \subcaptionbox{Step 6}{\includegraphics[trim={0 0 0 2.4cm},clip,width=5cm]{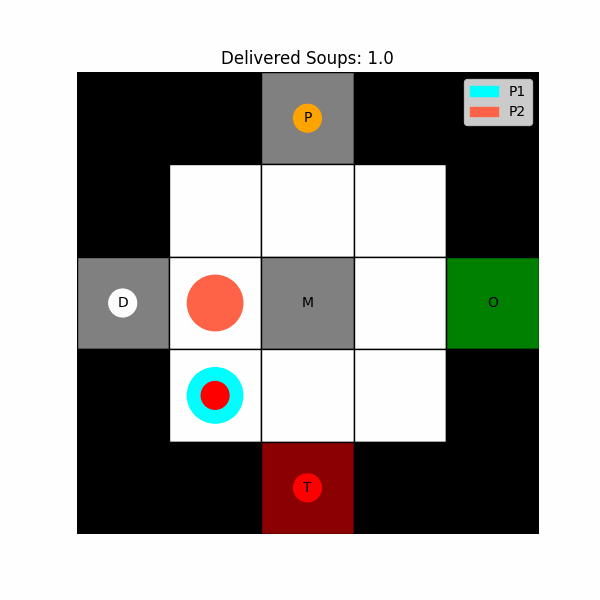}} &
        \subcaptionbox{Step 7}{\includegraphics[trim={0 0 0 2.4cm},clip,width=5cm]{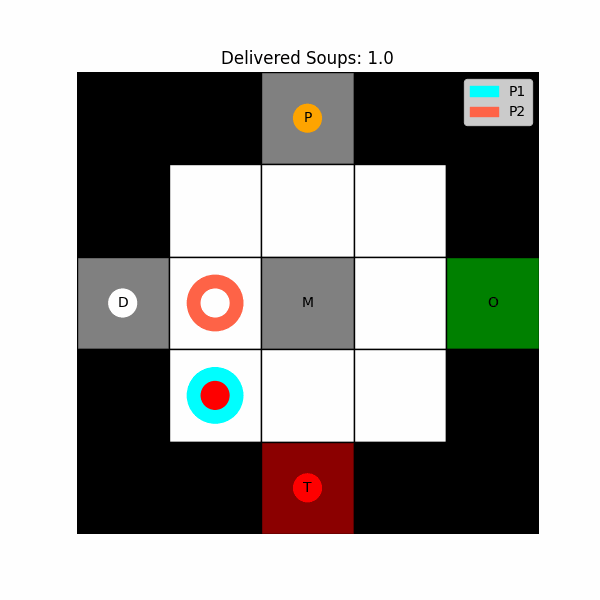}} &
        \subcaptionbox{Step 8}{\includegraphics[trim={0 0 0 2.4cm},clip,width=5cm]{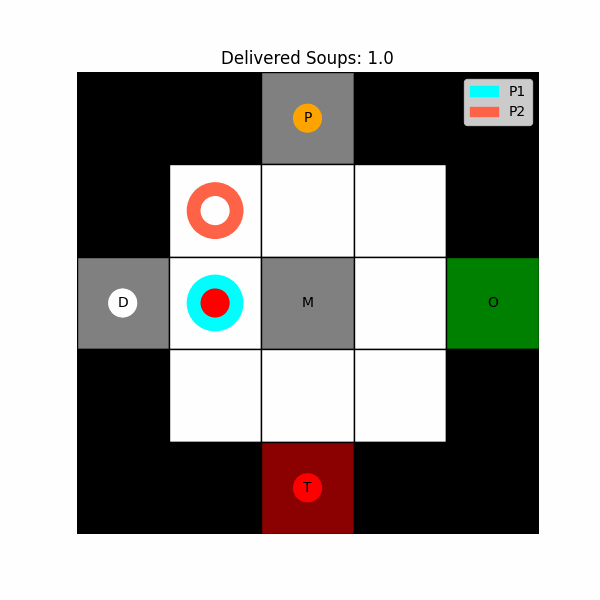}} &
        \subcaptionbox{Step 9}{\includegraphics[trim={0 0 0 2.4cm},clip,width=5cm]{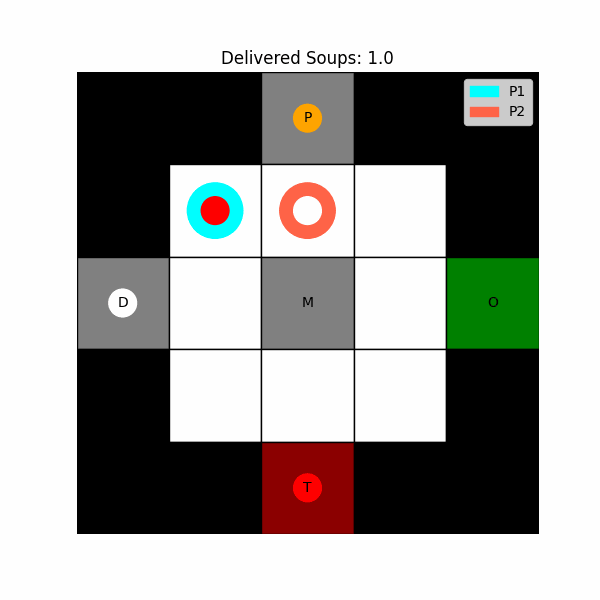}} \\
        \subcaptionbox{Step 10}{\includegraphics[trim={0 0 0 2.4cm},clip,width=5cm]{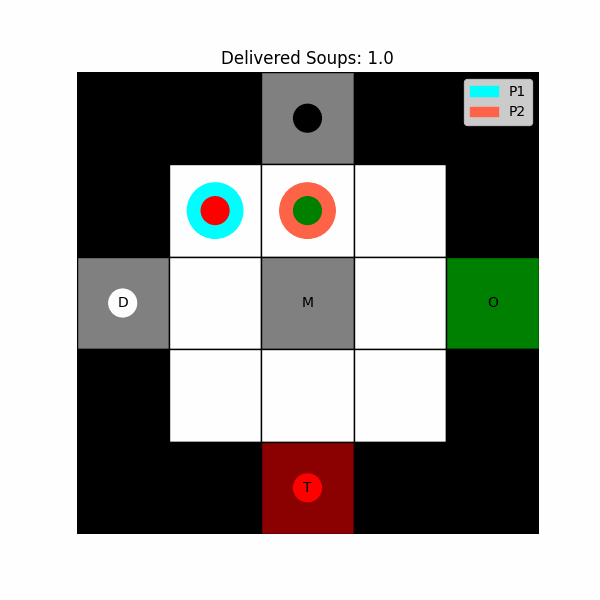}} &
        \subcaptionbox{Step 11}{\includegraphics[trim={0 0 0 2.4cm},clip,width=5cm]{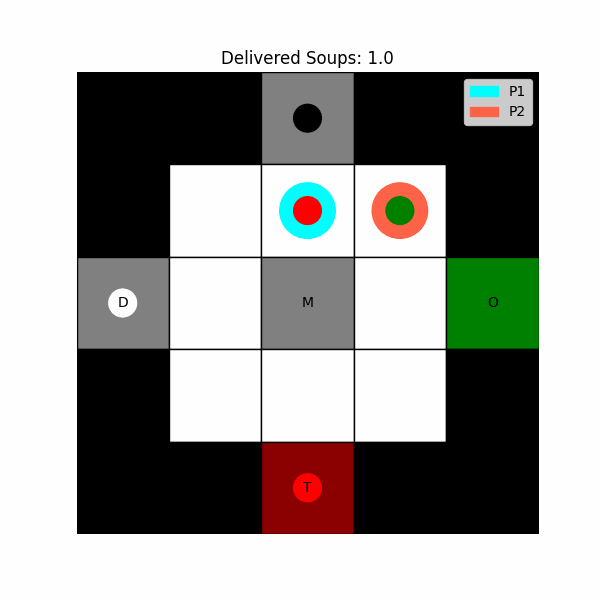}} &
        \subcaptionbox{Step 12}{\includegraphics[trim={0 0 0 2.4cm},clip,width=5cm]{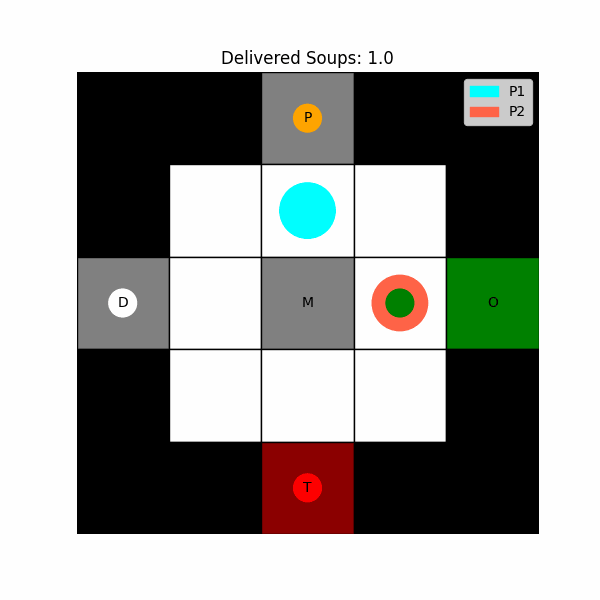}} &
        \subcaptionbox{Step 13}{\includegraphics[trim={0 0 0 2.4cm},clip,width=5cm]{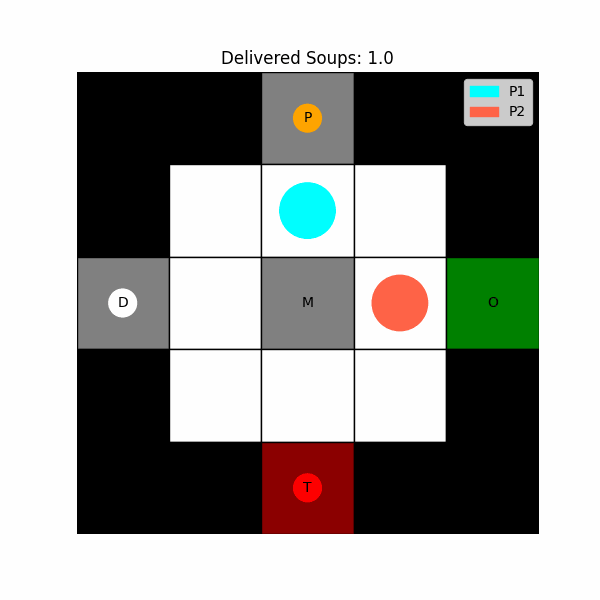}}
    \end{tabular}
    }
    \caption{Example trajectory extract in the collaborative cooking task.
    Blue and red circles represent the players. The red tile marks the tomato stand, the gray tile in the middle is a table, while the white circle indicates the plate stand. The top black/yellow circle shows the pot, with black representing empty and yellow representing filled. Finally, the green tile designates the delivery location where players must deliver their soup.
    (b) Blue player pick up a plate and a tomato. (c) Blue player passes its tomato on the table (d) Red player grabs the tomato on the table (e) Red player places tomato in pot (h) Red player gets a plate (k) Red player fills its plate with the soup (n) Red player delivers the soup.}
    \label{fig:trajectory}
\end{figure}

Our implementation of the cooking task is inspired from the Overcooked game, which is widely used as a benchmark for RL and IRL~\citep{carroll_utility_learning_about_2019, agapiou_melting_pot_2_2023, haynam2025multi}. We provide a visualisation of the sequence of actions needed to deliver a soup in Figure~\ref{fig:trajectory}. In our setup, game states are described by:

In this game, two players act as chefs, working together to prepare and deliver tomato soups. Players must collect tomatoes, cook them, carry dishes and deliver soups. The kitchen layout (see Figure~\ref{fig:layout}) requires object interaction in the correct sequence and tight coordination. While players cannot move through each other, they start at the same position to make the game invariant to player permutations.

\begin{enumerate}
    \item Player positions. Players cannot have the same position, except at the start where they overlap.
    \item Player state: carrying nothing, a plate, a tomato or a soup.
    \item Pot state: empty or ready.
\end{enumerate}
Notably, players do not have orientations, and have full observability of the game. A player has 5 actions:
\begin{enumerate}
    \item Moving up, down, left and right.
    \item Interact. The action's effect depends on the nearby tiles: take a plate, put tomatoes into the pot, etc.
\end{enumerate}
We chose the largest configuration that could fit into GPU RAM when performing Bayesian inference.

For all experiments, we used a discount factor of $\gamma=0.9$, and an entropy parameter of $\beta=0.05$.

\subsection{Implementation.} Our code is available at \href{https://anonymous.4open.science/r/altruism_mairl}{https://anonymous.4open.science/r/altruism\_mairl}.

\subsection{Compute.} $1 \times$ \texttt{INTEL(R) XEON(R) PLATINUM 8562Y+/NVIDIA L40S}.

\noindent DRP and QIG were both evaluated faster on CPU. See Table~\ref{tab:compute_time} for computation times per method in the random MG benchmark.

\begin{table}[t]
\centering
\caption{SGLD hyperparameters for DRP and PORP.\label{table:sgld_hypers}}
\begin{tabular}{lcc}
\toprule
\textbf{Method / Posterior} & \(\alpha\) & \(\epsilon_0\) \\
\midrule
DRP (sampling) & 0.05 & 0.1 \\
PORP (policy posterior) & 0 & 0.2 \\
PORP (reward posterior) & 0.5 & 1.5 (Random MG), $\;\;$ 5 (Overcooked)\\
\bottomrule
\end{tabular}
\end{table}

\begin{table}[t]
\centering
\caption{Priors, and hyperparameters used in experiments.}
\label{tab:priors}
\begin{tabular}{llc}
\toprule
\textbf{Parameter} &\textbf{Prior} & \textbf{Hyperparameters} \\
\midrule
$\beta$ & $\bel(\beta) = \mathrm{Exp}(l)$ & $l$ = 10 \\
$\theta_{s,a}$ (policy logits) & $\bel(\theta) = \mathcal{N}(0, \sigma^2)$ & N / A (Random MG), $\quad \sigma = \frac{1}{40}$ (Overcooked) \\
$\phi_{s,a}$ (intrinsic reward parameters)& $\bel(\phi) = \mathcal{N}(0, \sigma^2)$ & $\sigma = \frac{1}{40}$ (DRP), $\quad \frac{1}{6}$ (PORP) \\
$\lambda_i$ & $\bel(\lambda_i) = \mathcal{U}([-\lambda_\text{min}, \lambda_\text{max}])$ & $\lambda_\text{max} = - \lambda_\text{min} = 5$ \\
\bottomrule
\end{tabular}
\end{table}

\begin{table}[t]
    \centering
    \caption{Additional hyperparameters used by algorithms.}
    \label{tab:porp_hyperparameters}
    \begin{tabular}{rcc}
        \toprule
        \textbf{Hyperparameter} & \textbf{Random MGs} & \textbf{Overcooked} \\
        \hline
        $c$ (PSG) & \multicolumn{2}{c}{500}\\
        $c$ (NIG) & 50.000 & -- \\
        $\beta_\text{min}$ & 0.05 & 0.03 \\
        Batch size (MAMQL) & \multicolumn{2}{c}{2048}\\
        Buffer size (MAMQL) & \multicolumn{2}{c}{16.600}\\
        Batch size (MAAIRL) & 512 & -- \\
        Buffer size (MAAIRL) & 10.000 & -- \\
        \bottomrule
    \end{tabular}
\end{table}

\begin{table}[t]
    \centering
    \caption{Average computation time for all methods on the benchmarked random MGs (Figure~\ref{tab:benchmark_results}), evaluated on an \texttt{INTEL(R) XEON(R) PLATINUM 8562Y+/NVIDIA L40S}.}
    
    \label{tab:compute_time}
    \begin{tabular}{lc}
    \toprule
    \textbf{Method} & \textbf{Compute time} (s) \\
    \midrule
    PORP-PSG & 384 $\pm$ 2 \\
    PORP-QIG & 684 $\pm$ 4 \\
    DRP & 6200 $\pm$ 19  \\
    \hdashline
    MAMQL & 312 $\pm$ 1 \\
    MAAIRL & 2917 $\pm$ 11 \\
    \midrule
    \multicolumn{2}{l}{\textit{Without groups}} \\
    \midrule
    PORP-PSG & 296 $\pm$ 2 \\
    PORP-QIG & 277 $\pm$ 3 \\
    DRP & 900 $\pm$ 3 \\
    \bottomrule

\end{tabular}
\end{table}

\end{document}